\documentclass[aps,pra,superscriptaddress,longbibliography,twocolumn]{revtex4-2}

\usepackage{graphicx}
\usepackage{amsmath}
\usepackage{amssymb}
\usepackage{hyperref}
\usepackage[utf8]{inputenc}
\usepackage{mathtools}
\usepackage[english]{babel}
\usepackage{bbm}
\hypersetup{colorlinks=true, linkcolor=blue, citecolor=blue, urlcolor=blue}
\usepackage{xcolor}
\usepackage{braket}
\usepackage[normalem]{ulem}

\newcommand{\ie}{i.\,e.,\ }

\begin{document}
\title{Stabilization of dark states in emitter arrays coupled to a half-waveguide}

\author{Oriol Rubies-Bigorda}
\email{oriol.rubiesb@gmail.com}
\affiliation{Physics Department, Massachusetts Institute of Technology, Cambridge, Massachusetts 02139, USA}
\affiliation{Department of Physics, Harvard University, Cambridge, Massachusetts 02138, USA}
\author{Susanne F. Yelin}
\affiliation{Department of Physics, Harvard University, Cambridge, Massachusetts 02138, USA}
\author{Ana Asenjo-Garcia}
\affiliation{Department of Physics, Columbia University, New York, New York 10027, USA}
\author{Stuart J. Masson}
\affiliation{Department of Physics, University of South Florida, Tampa, Florida 33620, USA}

\begin{abstract}
The radiative properties of quantum emitters are profoundly influenced by their electromagnetic environment. When coupled to a waveguide terminated at one end by a mirror, distant emitters interact strongly via virtual photon exchange, leading to collective superradiant and subradiant states with enhanced or suppressed decay rates. We demonstrate that applying optimal frequency shifts to each emitter enables the formation of perfect single-excitation dark states (\ie states with zero decay rate) and near-perfect multi-excitation dark states in small ensembles. These collective states can be deterministically prepared with high fidelity using classical driving fields or few-photon pulses propagating along the waveguide. These results, readily implementable in superconducting qubit platforms, open new avenues for quantum information storage, networking and control of light. 
\end{abstract}

\maketitle


\section{Introduction}
\label{section: introduction}

Quantum light-matter interfaces enable the control of light at the few-photon level and form the backbone of quantum information processing. In ensembles coupled to a common electromagnetic environment, photon exchange mediates collective effects, giving rise to superradiant and subradiant states that respectively exhibit an enhanced and suppressed decay rate as compared to individual emitters~\cite{Dicke54,Gross82}. Subradiant states arise from a reduced coupling with the electromagnetic environment due to destructive interference of the electromagnetic fields radiated by each emitter. The ability to store information in extremely long-lived states offers a tantalizing promise for photon storage~\cite{Asenjo17PRX,Manzoni18,Ballantine21,Rubies22PRR}, metrology~\cite{Facchinetti18,Henriet19,Zafra26,Wang26} and quantum computation~\cite{Palma96,Zanardi97,Lidar98,Beige00,Paulisch16,Shah24,Rubies25}.

The promise of subradiant states is also their greatest limitation: decoupling from the electromagnetic field prevents direct excitation via that field. Consequently, a key challenge is to engineer subradiant states that have strongly suppressed emission while remaining addressable. In atomic clouds, subradiant states can be probabilistically populated by letting an excited ensemble decay~\cite{Guerin16,Solano17,Ferioli21PRX,Douglas26arxiv}. While the efficiency of this process can be optimized by appropriate choice of initial state~\cite{Holzinger22,Santos22,Rubies23PRA}, the process lacks control over the specific state produced. In subwavelength free-space arrays~\cite{Rui20,Douglas26arxiv}, subradiant spin waves can be selectively addressed by periodically modulating the emitter frequencies~\cite{Rubies22PRR}, via multi-photon transitions~\cite{Rusconi21}, or through an auxiliary excited emitter~\cite{Masson20PRR,Patti21,Brechtelsbauer21,CastellsGraells21,Fernandez22}. However, these proposals only work for large numbers of emitters where the spin wave picture is valid. 

Dark states with small emitter number can be produced for emitters coupled to a confined electromagnetic environment, such as a cavity or waveguide, by driving them externally to that environment~\cite{Zanner22,Zhenjie23,Du26arxiv}. This external drive is able to directly imprint the desired phase profile on the emitters. However, extending these methods to improve the lifetimes and controlled generation of subradiant states in systems with few emitters and multiple excitations remains an important challenge. In particular, applications of dark states require precise control over their properties and generation, optimally in arbitrary configurations.

In this work, we propose a method to prepare single- and multi-excitation dark states for small arrays of emitters coupled to a one-dimensional waveguide terminated at one end by a mirror. Unlike in the conventional permutationally symmetric limit of all-to-all dissipative interactions, emitters in a half-waveguide experience spatially varying decay rates and coherent exchange interactions, leading to a rich interplay between dissipative and coherent dynamics. We show that this interplay can be harnessed to perfectly stabilize arbitrary single-excitation dark states and manifolds using relative detunings between emitters. Systems composed of multiple such subsystems, separated in frequency to reduce crosstalk, thus play host to near-perfect multi-excitation dark states. These states can be deterministically prepared with weak classical driving fields or few-photon pulses. Our work outlines a strategy for multi-photon storage and release for quantum memories and quantum networks.

\begin{figure}
    \centering
    \includegraphics[width = \columnwidth]{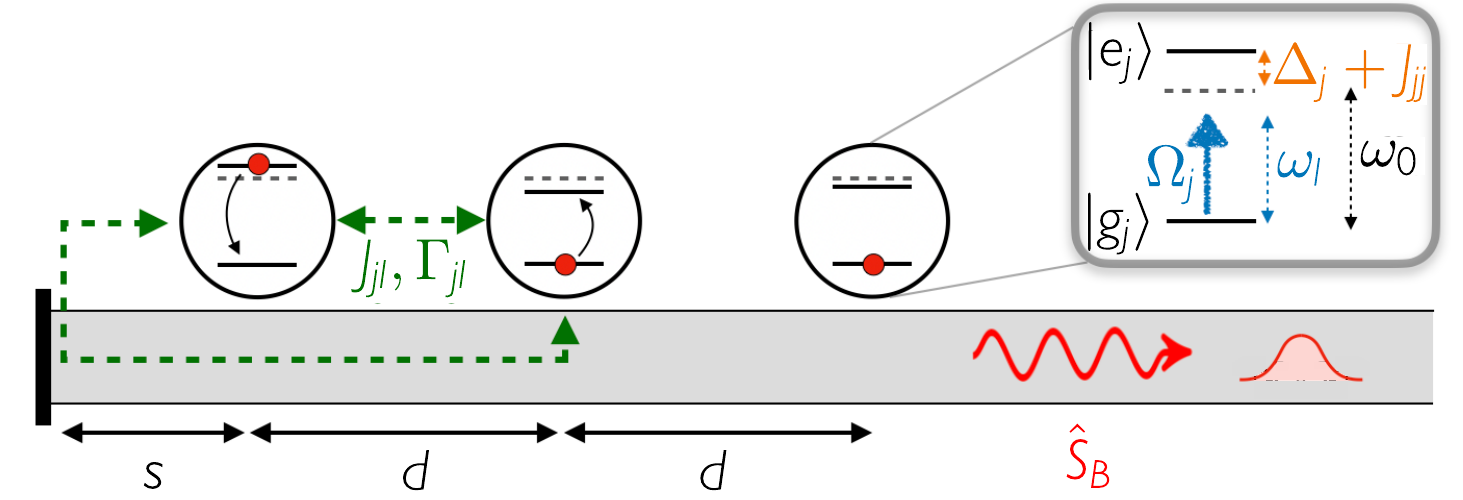}
    \caption{Array of $N$ two-level emitters coupled to a one-dimensional waveguide terminated by a mirror. The separation between the first emitter and the mirror is $s$, while the distance between emitters is $d$. The photonic modes of the waveguide induce a coherent $J_{jl}$ and dissipative $\Gamma_{jl}$ interaction between emitters $j$ and $l$, which can be described as the superposition of the two waves traveling from one emitter to the other (green dashed arrows). Photons are emitted by the array via action of the bright jump operator $\hat{S}_B$ with decay rate $\Gamma_B$. Additionally, each individual emitter can be detuned from its bare transition frequency by an amount $\Delta_j$ and driven at frequency $\omega_l$ at a Rabi frequency $\Omega_j$. }
    \label{fig: setup}
\end{figure}

\section{Model}
\label{section: model}

We consider $N$ two-level emitters (or qubits) of resonance frequency $\omega_0$ coupled to a one-dimensional photonic waveguide terminated at one end by a mirror, as illustrated in Fig.~\ref{fig: setup}. We consider the distances between the emitters and the mirror to be sufficiently small that the photon travel time between them remains negligible compared to the characteristic timescale of the emitters, allowing the photonic degrees of freedom to be traced over under a Markov approximation~\cite{Dung02}. The master equation that governs the qubits' dynamics then reads ($\hbar=1$)~\cite{Fayard21,Rubies25}
\begin{equation}
\label{eq: Master_equation}
    \frac{d\hat{\rho}}{dt} = -i \left[ \hat{H}, \hat{\rho}\right] + \mathcal{L}(\hat{\rho}).
\end{equation}
The Hamiltonian $\hat{H}$ and the Lindbladian $\mathcal{L}(\hat{\rho})$ respectively describe the coherent and dissipative interactions mediated by the waveguide modes
\begin{align}
\label{eq: Hamiltonian}
    \hat{H} &= \sum_{j} (\omega_0 + \Delta_{j} )  \hat{\sigma}_j^{eg} \hat{\sigma}_j^{ge} + \sum_{j,l} J_{jl}  \hat{\sigma}_j^{eg} \hat{\sigma}_l^{ge}, \\
\label{eq: Lindbladian}
    \mathcal{L}(\hat{\rho}) &= \sum_{j,l} \frac{\Gamma_{jl}}{2} \left( 2 \hat{\sigma}_l^{ge} \hat{\rho} \hat{\sigma}_j^{eg} - \hat{\sigma}_j^{eg} \hat{\sigma}_l^{ge} \hat{\rho} - \hat{\rho} \hat{\sigma}_j^{eg} \hat{\sigma}_l^{ge} \right),
\end{align}
where $\hat{\sigma}_j^{eg} = |e_j \rangle \langle g_j |$ and $\hat{\sigma}_j^{ge} = |g_j \rangle \langle e_j |$ are the raising and lowering operators for the $j$th qubit, located at position $x_j$. Each qubit can be detuned from the resonance frequency by $\Delta_j$, and each pair of qubits interacts with coherent $J_{jl}$ and dissipative $\Gamma_{jl}$ coupling strengths set by
\begin{equation}
\label{eq: coh_dis_int}
    J_{jl} - i \frac{\Gamma_{jl}}{2} = - i\frac{ \gamma_0}{2} \left( e^{i k_0 |x_j - x_l|} - e^{i k_0 |x_j + x_l|} \right).
\end{equation}
Here, $k_0 = 2 \pi/\lambda_0$, with $\lambda_0$ the resonant wavelength of the qubits, and $\gamma_0$ describes the coupling of an emitter to a full one-dimensional waveguide (\ie not terminated by a mirror). 

Interactions are mediated by photons traversing two paths, as illustrated by the green dashed arrows in Fig.~\ref{fig: setup}. The first contribution corresponds to a photon propagating directly from emitter $j$ to emitter $l$ with acquired phase proportional to their separation, $k_0 |x_j - x_l|$, as in a waveguide without a mirror. The second contribution arises from a photon that is reflected by the mirror with acquired propagation phase proportional to $k_0 |x_j + x_l| + \pi$, where the additional $\pi$ is due to the reflection. Due to the interference between both contributions, the decay rate of a single emitter to the waveguide, $\Gamma_{jj} = 2 \gamma_0 \sin^2 (k_0 x_j)$, depends on its position $x_j$~\cite{Wiegand20}. For qubits located at positions $x_j = m \lambda_0/2$ with $m \in \mathbb{N}$, we obtain perfect destructive interference, i.e., $\Gamma_{jj} = 0$. On the other hand, perfect constructive interference and maximal decay rate $2\gamma_0$ is attained at positions $x_j =\lambda_0/4+ m \lambda_0/2$ with $m \in \mathbb{N}$.

Emission occurs only towards the transmitting side of the waveguide, as indicated by the red arrow in Fig.~\ref{fig: setup}, and so the system’s Lindbladian can be expressed in terms of a single collective jump operator $\hat{S}_B$ with decay rate $\Gamma_B$, 
\begin{subequations}
\begin{align}
\label{eq: jump_operator}
    \hat{S}_B &=\sqrt{\frac{2 \gamma_0}{\Gamma_B}} \sum_j \sin (k_0 x_j) \hat{\sigma}_j^{ge}, \\
\label{eq: bright_rate}
    \Gamma_B &= \sum_j \Gamma_{jj} = 2 \gamma_0 \sum_j \sin^2 (k_0 x_j).
\end{align} 
\end{subequations}
The contribution of each emitter to the bright decay mode has strength proportional to its own position-dependent decay rate $\Gamma_{jj}$. Using the collective jump operator we can rewrite the master equation~(\ref{eq: Master_equation}) as
\begin{equation}
\frac{d\hat{\rho}}{dt} = -i \left( \hat{H}_\mathrm{nH}\hat{\rho} - \hat{\rho} \hat{H}_\mathrm{nH}^\dagger \right) + \Gamma_{B} \hat{S}_B \hat{\rho} \hat{S}_B^{\dagger},
\end{equation}
where we have defined the non-Hermitian Hamiltonian, $\hat{H}_\mathrm{nH} = \hat{H} - i \hat{H}_\mathrm{dis}$ and $\hat{H}_\mathrm{dis} = \Gamma_B \hat{S}_B^{\dagger} \hat{S}_B / 2$.

For the remainder of this work, we focus on evenly spaced chains of emitters, which can be fully characterized by the distance $s$ of the first emitter from the mirror and the spacing $d$ between emitters [see Fig.~\ref{fig: setup}]. Due to the periodicity of the coherent and dissipative couplings mediated by the waveguide field, arrays with $s+ n\lambda_0/2$ and $d+m\lambda_0$ are equivalent for all $n,m \in \mathbb{N}$.

\section{Lindbladian spectrum}
\label{section: dissipative interactions}

Collective interactions give rise to a rich landscape of subradiant and dark states. To gain insight into the nature of dark states in the system, we begin by diagonalizing the dissipative Hamiltonian, $\hat{H}_\mathrm{dis}$. Collective states that fulfill
\begin{equation}
\label{eq: Condition_Lindblaidan_darkstate}
    \hat{H}_{\mathrm{dis}} |\psi \rangle = \hat{S}_B |\psi \rangle = 0,
\end{equation}
do not emit into the waveguide under the action of the dissipative interactions, and we refer to them as \textit{Lindbladian dark states}. However, Eq.~(\ref{eq: Condition_Lindblaidan_darkstate}) does not ensure that $|\psi \rangle$ remains invariant under the evolution governed by the master equation~(\ref{eq: Master_equation}) because the Hamiltonian can couple $| \psi \rangle$ to other states, potentially leading to decay through interactions with bright states. A \textit{total dark state} is thus a simultaneous eigenstate of the coherent and dissipative interactions, \ie
\begin{equation}
\label{eq: eigenstate_nonHermitian_general}
    \hat{H}_{nH} |\phi \rangle = \left(\hat{H} - i \hat{H}_\mathrm{dis} \right) |\phi \rangle  = \left( E_\phi - i \frac{\gamma_\phi}{2} \right) |\phi\rangle.
\end{equation}
with $\gamma_\phi = 0$. In what follows, we discuss the existence of Lindbladian and total dark states in arrays of emitters coupled to a half-waveguide.

\subsection{Permutational symmetry, $d=m\lambda_0/2$ \label{subsection: permutational}}

\begin{figure}
    \centering
    \includegraphics[width = \columnwidth]{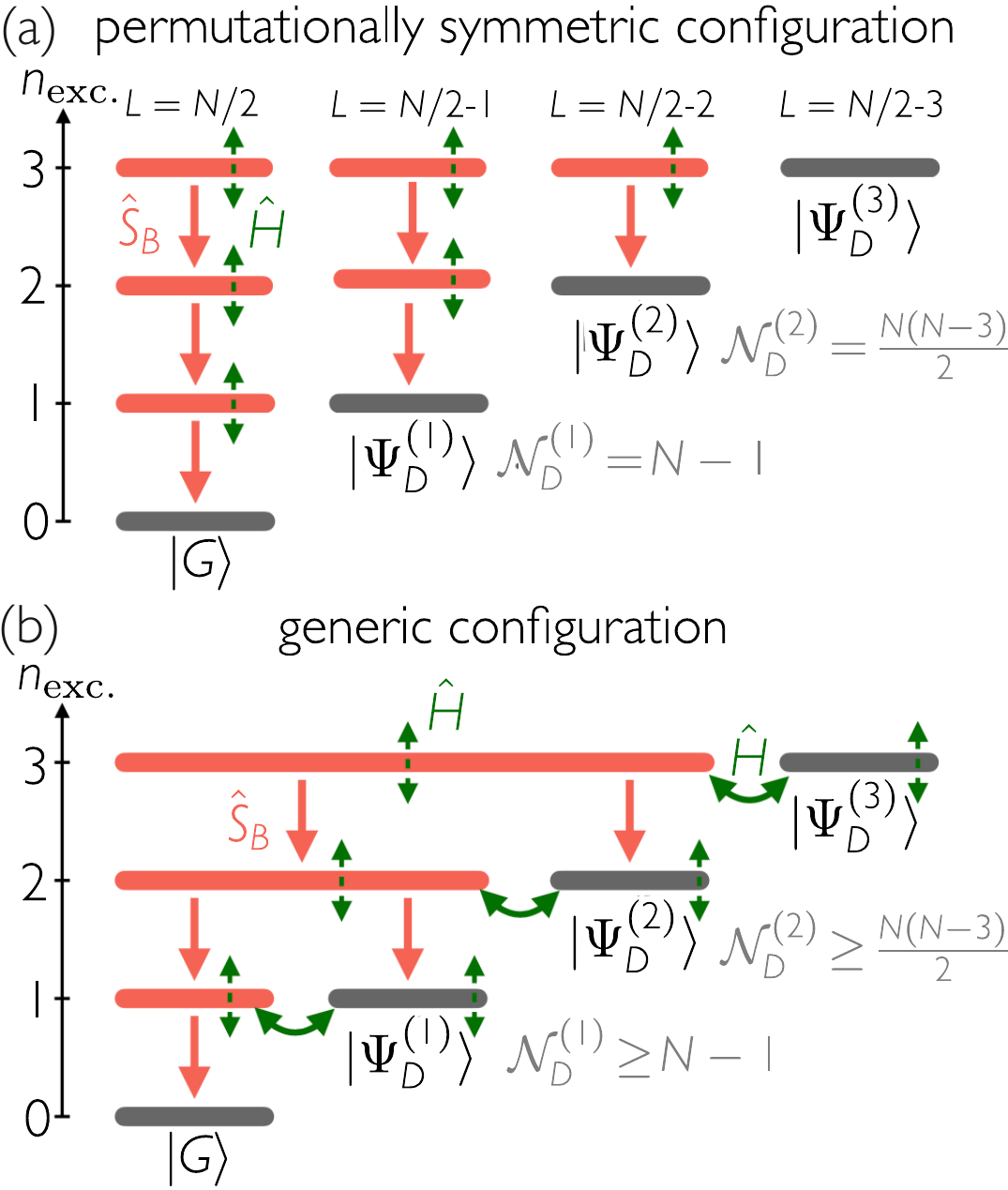}
    \caption{ Lindbladian spectrum or eigenstates of the dissipative Hamiltonian for (a) the permutationally symmetric configuration ($d=m \lambda_0/2$ with $m \in \mathbb{N}$) and (b) any other general configuration, for the four lowest excitation numbers $n_\mathrm{exc}$. In (a), decay (red arrows) only occurs within each ladder with well-defined total angular momentum L. The coherent Hamiltonian $\hat{H}$ gives rise to energy shifts of each state (green dashed double arrows). The end of each ladder corresponds to a set of $\mathcal{N}_\mathcal{D}^{(n_\mathrm{exc})}$ Lindbladian dark states, $|\psi_\mathcal{D}^{(n_\mathrm{exc})} \rangle$, colored in grey. (b) For general configurations $d \neq m \lambda_0/2$, photon emission does not conserve the total angular momentum of the system. Additionally, the coherent Hamiltonian $\hat{H}$ results in coupling between different Lindbladian eigenstates, thereby opening a path for Lindbladian dark states to decay via Lindbladian bright states (green double arrows). }
    \label{fig: spectrum}
\end{figure}

All emitters interact with equal strength if the distance between them is an integer or half-integer number of wavelengths, $d=m \lambda_0/2$ with $m \in \mathbb{N}$,
\begin{subequations}
\begin{align}
    \Gamma_{jl} & = 2 \gamma_0 \sin^2(k_0 s) (-1)^{m(j+l)}, \\
    J_{jl}& = - \gamma_0 \sin (k_0 s) \cos (k_0 s) (-1)^{m(j+l)}.
\end{align}
\end{subequations}
The jump operator takes the symmetrical form \begin{equation}
\hat{S}_B = \sum_{j=0}^{N-1} (-1)^{mj} \hat{\sigma}_j^{ge} / \sqrt{N}
\end{equation}
and the collective decay rate is $\Gamma_B = 2 N \gamma_0 \sin^2(k_0 s)$. Dissipative and coherent interactions share the same eigenstates, as the non-Hermitian Hamiltonian can be expressed as
\begin{equation}\hat{H}_\mathrm{nH} = (J_B - i \Gamma_B/2) \hat{S}_B^\dagger \hat{S}_B,    
\end{equation}
with $J_B = -N \gamma_0 \sin (k_0 s) \cos (k_0 s)$. Due to the permutational symmetry, the eigenstates of the non-Hermitian Hamiltonian can be expressed as states of a single collective angular spin~\cite{Gross82}, where photon emission cannot alter the total angular momentum $L$ of the system. As such, decay occurs only within each ladder of fixed $L$ until the total dark states at the bottom of each ladder are populated, as shown in Fig.~\ref{fig: spectrum}(a). For $s = \lambda_0/4$, $J_B = 0$ and the system reduces to the textbook case of Dicke superradiance. For $s \neq \lambda_0/4$, the eigenstates of the system additionally acquire trivial energy shifts, illustrated by the green dashed arrows in Fig.~\ref{fig: spectrum}(a).

\subsection{General configuration, $d\neq m\lambda_0/2$ \label{subsubsection: Lindbladuan_generalconfiguration}}

In the absence of permutational symmetry, $d \neq m \lambda_0/2$,  the contribution of each emitter to the bright jump operator varies and the collective spin picture described above cannot be used. 
We can diagonalize the dissipative interactions $\hat{H}_\mathrm{dis}$ and find that, for the manifold with $n_{\mathrm{exc}}$ excitations, there exist
\begin{equation}\mathcal{N}_\mathcal{D}^{n_\mathrm{exc}} \geq {N\choose n_\mathrm{exc}} - {N\choose n_\mathrm{exc}-1}
\end{equation}
Lindbladian dark states which satisfy Eq.~(\ref{eq: Condition_Lindblaidan_darkstate}) [see Appendix~\ref{appendix: numberDarkStates} for details]. Note that the number of dark states with a given number of excitations can be larger than that of the permutationally symmetric configuration (for which the equality holds) if some emitters are dissipatively decoupled from the half-waveguide ($\Gamma_{jj} =0$).

Photon emission from a bright eigenstate does not generate another eigenstate of $\hat{H}_\mathrm{dis}$ [see Appendix~\ref{appendix: proof_jump}]. Repeated action of the jump operator therefore populates Lindbladian dark states for general initial states (including the fully excited state). For general configurations, the coherent and dissipative Hamiltonians do not commute,  $[\hat{H},\hat{H}_\mathrm{dis}] \neq 0$ [see Appendix~\ref{appendix: commutation}]. As a result, the Lindbladian dark states acquire both energy shifts and coupling to other states [see Fig.~\ref{fig: spectrum}(b)], and are in general not total dark states.

\section{Stabilization of dark states}
\label{section: dark_state_engineering}

Decay from Lindbladian dark states occurs due to coupling to bright states. For general configurations, the decay rate of the darkest eigenstate of the non-Hermitian Hamiltonian scales as $N^{-3}$ as in the full waveguide case~\cite{Asenjo17PRX,Zhang19}, resulting in significant decay for small emitter arrays. In this section, we introduce a method to achieve perfect single-excitation dark states and near-perfect multi-excitation dark states even for small emitter numbers, by applying tailored frequency shifts to each two-level emitter.

\subsection{Single-excitation dark states}
\label{subsection: 1excitationDark}

\begin{figure}
    \includegraphics[width=\columnwidth]{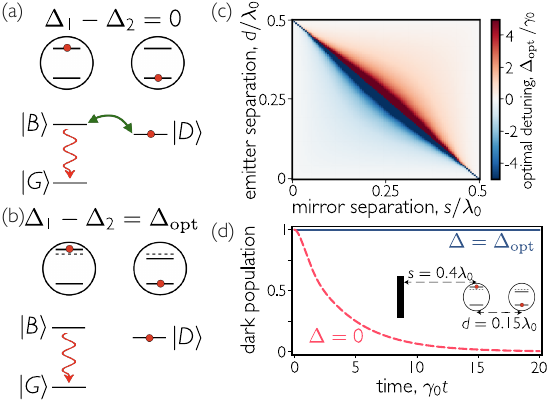}
    \caption{(a) For two emitters at a distance $d \neq m \lambda_0/2$ and in the absence of a relative detuning ($\Delta_1-\Delta_2=0$), the bright $|B\rangle$ and dark $|D\rangle$ Lindbladian eigenstates are coupled via the coherent interactions $\hat{H}$, providing a decay path for $|D\rangle$. (b) Applying the optimal relative detuning $\Delta_1-\Delta_2=\Delta_\mathrm{opt}$, the coupling is canceled, and $|D\rangle$ becomes a total dark state of the full master equation. (c) Optimal detuning $\Delta_\mathrm{opt}$ as a function of the separation $s$ of the first emitter from the mirror and the distance $d$ between emitters. (d) Overlap of the density matrix $\hat{\rho}(t)$ with the dark state $|D\rangle$ over time for a two-emitter array with $s=\lambda_0/4$ and $d=0.15\lambda_0$ initialized at $|D\rangle$ with (solid blue) optimized detuning and (dashed red) $\Delta=0$. }
    \label{fig: engineer_dark_2atoms}
\end{figure}

We first consider the case of two emitters. Alongside the ground state $|G \rangle \equiv |gg\rangle$ and fully excited state $|E \rangle \equiv |ee\rangle$, the eigenstates of the dissipative interactions are the single-excitation states
\begin{align}
\label{eq: 2atoms_brightstate}
    |B\rangle &= \hat{S}_B^\dagger |G\rangle =\frac{\sin(k_0 x_1) |eg\rangle + \sin(k_0 x_2) |ge\rangle}{\sqrt{\sin^2(k_0 x_1) + \sin^2(k_0 x_2)}}, \\
\label{eq: 2atoms_darkstate}
    |D\rangle &= \frac{\sin(k_0 x_2) |eg\rangle - \sin(k_0 x_1) |ge\rangle}{\sqrt{\sin^2(k_0 x_1) + \sin^2(k_0 x_2)}}.
\end{align}
The former satisfies $\hat{H}_\mathrm{dis} |B \rangle = (\Gamma_B / 2) |B\rangle$ and decays to the ground state by emitting a photon into the waveguide at a rate $\Gamma_B$. The latter fulfills $\hat{H}_{\mathrm{dis}} |D\rangle = 0$ and is a single-excitation Lindbladian dark state. However, it is not an eigenstate of the coherent Hamiltonian and can therefore decay through the bright state $|B\rangle$ [see Fig.~\ref{fig: engineer_dark_2atoms}(a)].  

An additional coupling between $|D\rangle$ and $|B\rangle$ can be induced by applying a relative detuning $\Delta_1  -\Delta_2$. As illustrated in Fig.~\ref{fig: engineer_dark_2atoms}, this coupling can cancel out the intrinsic coupling mediated by the half-waveguide, which renders $|D\rangle$ a total dark state. If both emitters are bright ($\Gamma_{jj} \neq 0$ for $j \in \{ 1,2 \}$), the optimal detuning is
\begin{align}
\label{eq: optimal_detuning_2atoms}
    \frac{\Delta_\mathrm{opt}}{\gamma_0} =\sin(k_0 x_1) \left( \cos (k_0 x_1) - \cos (k_0 x_2) \frac{\sin(k_0 x_1)}{\sin(k_0 x_2)}\right) ,
\end{align}
and is plotted in Fig.~\ref{fig: engineer_dark_2atoms}(c).
Note that $\Delta_\mathrm{opt}$ diverges when the second emitter is positioned at a node of the waveguide field, where $\Gamma_{22}=0$. This occurs for $x_2=\lambda_0 n/2$, or equivalently $d=\lambda_0 n/2-s$, with $n\in\mathbb{N}$ [the dark diagonal in Fig.~\ref{fig: engineer_dark_2atoms}(c)]. In this configuration, the dark state becomes unentangled and localized on the second emitter, while the finite coherent interaction $J_{12}$ between the emitters cannot be compensated by local detunings. By contrast, no divergence occurs when the first emitter is positioned at a node, such that $\Gamma_{11}=0$ ($x_1=s=\lambda_0 n/2$, with $n\in\mathbb{N}$), because the coherent interaction between the emitters then vanishes.

We can generalize the treatment described above to stabilize arbitrary single-excitation dark states for $N$ emitters. The dissipative interactions $\hat{H}_\mathrm{dis}$ only support one single-excitation bright eigenstate, $|\psi_B^{(1)} \rangle = \hat{S}_B^\dagger |G\rangle$, where $|G\rangle = \ket{g}^{\otimes N}$ is the ground state. The remaining $N-1$ single-excitation states are Lindbladian dark states, which can be written as
\begin{equation}
\label{eq: 1exc_dark_Natoms}
|\psi_\mathcal{D}^{(1)} \rangle = \sum_{j=1}^N \alpha_j  \hat{\sigma}_j^{eg} |G\rangle
\end{equation}
with $\sum_j \alpha_j \sin (k_0 x_j) = 0$ and $\sum_j |\alpha_j|^2 = 1$.

A Lindbladian eigenstate $|\psi\rangle = \sum_j \psi_j \hat{\sigma}_j^{eg} |G\rangle$ is also an eigenstate of the full dynamics if it fulfills Eq.~(\ref{eq: eigenstate_nonHermitian_general}), \ie if $\hat{H} |\psi \rangle = E_\psi |\psi \rangle $. This results in a linear set of $N$ equations with $N$ unknowns: the energy shift $E_\psi$ and $N-1$ relative detunings. Setting without loss of generality $\Delta_1=0$, its solution is
\begin{equation}
\label{eq: optimaldet_Natoms_1exc}
    \begin{pmatrix}
    \Delta_2 \\ \vdots \\ \Delta_N \\ E_\psi 
    \end{pmatrix}
    = - \mathbf{M}^{-1} \mathbf{J} 
    \begin{pmatrix}
    \psi_1 \\ \vdots \\ \psi_N
    \end{pmatrix},
\end{equation}
where $\mathbf{J}$ is an $N\times N$ matrix with elements $J_{jl}$ and $\mathbf{M}$ is the $N\times N$ matrix 
\begin{equation}
   \mathbf{M} = \begin{pmatrix}
       0 &0&\dots&0&-\psi_1 \\
       \psi_2&0&\dots&0&-\psi_2 \\
       0&\psi_3&\dots&0&-\psi_3 \\
       \vdots&\vdots&\ddots&\vdots&\vdots \\
       0&0&\dots&\psi_N&-\psi_N
   \end{pmatrix} .
\end{equation}

This method allows us to perfectly stabilize a target dark state $|\psi_\mathcal{D}^{(1)} \rangle$. Generically, all other Lindbladian dark states are still coupled to the bright state and so decay. If we instead choose the bright state $| \psi_B^{(1)} \rangle$ as an eigenstate of the Hamiltonian, the perpendicular subspace spanned by the $N-1$ Lindbladian dark states becomes perfectly dark. However, the eigenstates of this subspace are automatically fixed by the specific choice of detunings and may exhibit different energy shifts. As a result, a general target Lindbladian dark state will undergo a non-trivial unitary evolution within the dark subspace.

\subsection{Multi-excitation dark states}
\label{subsection: engineeringdark_multi}

The existence of a single-excitation dark state does not guarantee the existence of a multi-excitation dark state due to the nonlinearity of two-level emitters. This occurs because the commutation of the coherent $\hat{H}$ and dissipative $\hat{H}_\mathrm{dis}$ Hamiltonians beyond a single excitation becomes increasingly stringent, and can only be attained for the permutationally symmetric configuration, as proven in Appendix~\ref{appendix: commutation}. To stabilize an arbitrary Lindbladian dark state with more than one excitation, one needs to solve for the detuning pattern that satisfies Eq.~(\ref{eq: eigenstate_nonHermitian_general}) with zero decay rate $\gamma_\phi = 0$. This results in an overdetermined set of ${N\choose n_\mathrm{exc}}$ linear equations but only $N$ free parameters. For the permutationally symmetric configurations, the system of equations is trivially solved by $\Delta_j = 0$ due to the commutation of coherent and dissipative interactions. For other configurations, the system of equations cannot be solved exactly and perfect multi-excitation total dark states do not exist. Instead, Lindbladian dark states can be made deeply subradiant via a numerical optimization of the detunings [see Appendix~\ref{Appendix: 4atoms_2exc}].

Alternatively, we can identify a class of near-perfect total dark states with $n_\mathrm{exc}$ that does not require numerical optimization. The system is divided in $n_\mathrm{exc}$ subsystems containing at least two bright emitters each. The jump operator can be written as a sum over subsystems, $\hat{S}_B = \sum_{b=1}^{n_\mathrm{exc}} \hat{S}_b $. If each subsystem is in a Lindbladian dark state of $\hat{S}_b$, then their product state is a dark state of $\hat{S}_B$. Separating the subsystems in frequency space strongly suppresses interactions between them, and so application of the optimal single-excitation detuning pattern described in Sec.~\ref{subsection: 1excitationDark} to each subsystem creates near-perfect total dark states.  

\begin{figure}
    \includegraphics[width=\columnwidth]{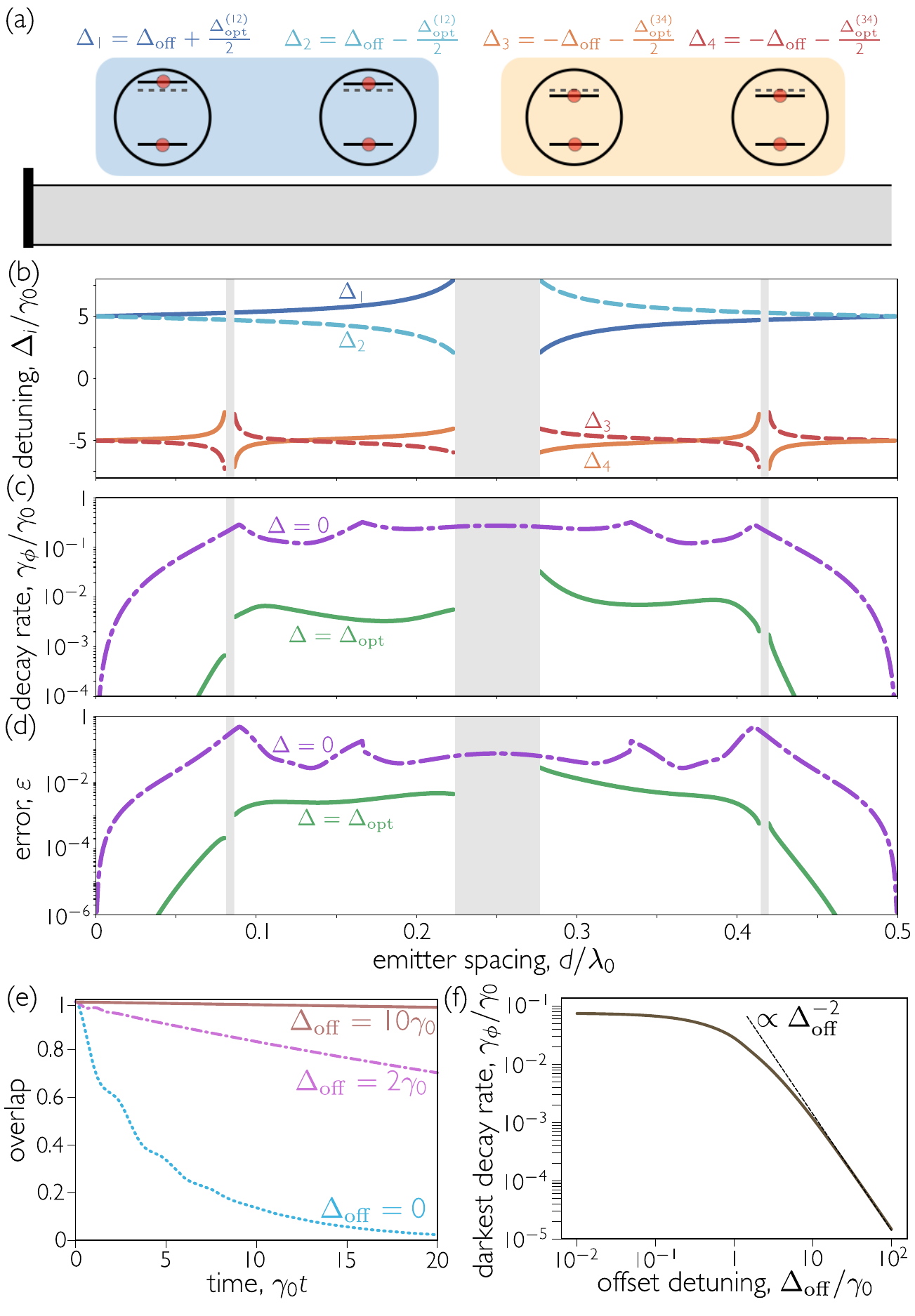}
    \caption{(a) Separable two-excitation dark state $|D_{12,34} \rangle = |D_{12} \rangle \otimes |D_{34} \rangle$, consisting of the product of two single-excitation dark states: $|D_{12} \rangle$ corresponds to the subsystem formed by the first two emitters (shaded blue), and $|D_{34} \rangle$ corresponds to the subsystem comprised of the third and fourth emitters. A relative detuning $\pm \Delta_\mathrm{off}$ is applied to the subsystems on top of the stabilizing detuning $\Delta_{\mathrm{opt}}$. (b) Optimal detuning $\Delta_j$ of each emitter $j$ as a function of $d$ for $s=\lambda_0/4$ with $ \Delta_\mathrm{off} = 5 \gamma_0$. (c) Decay rate $\gamma_\phi$ of the darkest two-excitation eigenstate, $\ket{\phi_D}$, of $\hat{H}_{nH}$ for zero (dot-dashed purple curve) and optimal (green curve) detuning. (d) One minus overlap of the separable two-excitation dark state $|D_{12,34} \rangle$ with the darkest eigenstate $| \phi_D \rangle$. In (b-d), the gray regions indicate parameter regimes where the optimal detuning grows without bound. (e) Overlap of $\hat{\rho}(t)$ with $| \phi_D \rangle$ during the decay process for various $\Delta_\mathrm{off}$ for $s=\lambda_0/4$ and $d=0.15\lambda_0$. (f) Darkest decay rate $\gamma_\phi$ as a function of $\Delta_\mathrm{off}$ for $s=\lambda_0/4$ and $d=0.15\lambda_0$. For $\Delta_\mathrm{off} \geq \gamma_0$, the decay rate scales as $\gamma_\phi \propto \Delta_\mathrm{off}^{-2}$.}
    \label{fig: engineer_Dark}
\end{figure}

We consider the smallest system size that supports two-excitation Lindbladian dark states, $N=4$ bright emitters, as an illustrative example of the protocol. We divide the array into two subsystems that respectively comprise the first two emitters and the last two emitters, as shown in Fig.~\ref{fig: engineer_Dark}(a), and introduce the two-excitation separable Lindbladian dark state
\begin{align}
\label{eq: darkstate_2exc_separable}
    |D_{12,34} \rangle &= |D_{12} \rangle \otimes |D_{34} \rangle \nonumber \\
    &= \frac{\sin (k_0 x_2) |eg\rangle - \sin (k_0 x_1) |ge\rangle}{\sqrt{\sin (k_0 x_1)^2 + \sin (k_0 x_2)^2}} \nonumber \\ &\quad \otimes \frac{\sin (k_0 x_4) |eg\rangle - \sin (k_0 x_3) |ge\rangle}{\sqrt{\sin (k_0 x_3)^2 + \sin (k_0 x_4)^2}}
\end{align}
as the state we wish to stabilize. Similarly, we can define the states $|D_{13,24} \rangle$ and $|D_{14,23} \rangle$. The two-excitation state $|D_{12,34} \rangle$ is near-perfectly stabilized by the optimal detunings
\begin{subequations}
\begin{align}
\label{eq: optimaldetunings_4at_2exc}
\Delta_1 = \Delta_\mathrm{off} + \frac{\Delta_\mathrm{opt}^{(12)}}{2}, \quad & \quad  
\Delta_2 = \Delta_\mathrm{off} - \frac{\Delta_\mathrm{opt}^{(12)}}{2}, \\
\Delta_3 = -\Delta_\mathrm{off} + \frac{\Delta_\mathrm{opt}^{(34)}}{2}, \quad & \quad
\Delta_4 = -\Delta_\mathrm{off} - \frac{\Delta_\mathrm{opt}^{(34)}}{2}.
\end{align}
\end{subequations}
Here, $2\Delta_\mathrm{off}$ corresponds to a global detuning between both subsystems, and $\Delta_\mathrm{opt}^{(12)}$ and $\Delta_\mathrm{opt}^{(34)}$ respectively denote the optimal detuning of the single-excitation dark state within each subsystem.

Frequency separation of the subsystems can be used to stabilize $\ket{D_{12,34}}$ in a wide range of configurations. Using the optimal detunings [Fig.~\ref{fig: engineer_Dark}(b)], we find that the darkest two-excitation eigenstate $\ket{\phi_D}$ of the non-Hermitian Hamiltonian has a strongly suppressed decay rate $\gamma_\phi \propto \Delta_{\mathrm{off}}^{-2}$ as compared to the case of zero detuning [Fig.~\ref{fig: engineer_Dark}(c)] and near-unit overlap with our target product dark state, $1 - \mathcal{F} = 1- |\langle D_{12,34} | \phi_D \rangle|^2 \propto \Delta_{\mathrm{off}}^{-2}$ [Fig.~\ref{fig: engineer_Dark}(d)]. As shown in Fig.~\ref{fig: engineer_Dark}(e), even a modest detuning between the two subsystems significantly lengthens the lifetime of the two-excitation dark state. We note that stabilization of $\ket{D_{12,34}}$ requires $\Delta_{\mathrm{off}} \gg \Delta_\mathrm{opt}^{(12)},\Delta_\mathrm{opt}^{(34)}$, and thus requires greater frequency separation in configurations with divergent optimized detunings. Furthermore, the generalization to larger excitation numbers requires more subsystems all separated in frequency space.

A perfect two-excitation dark state cannot be achieved in arbitrary configurations because the number of free parameters when solving Eq.~(\ref{eq: eigenstate_nonHermitian_general}) with $\gamma_\phi=0$ is too small. However, the number of free parameters can be increased by adding additional terms to the Hamiltonian. This could be done via direct interactions independent of the waveguide, e.g., capacitive interactions in superconducting circuits~\cite{Kannan23,Zanner22} or Rydberg interactions for atoms~\cite{Ocola24}. For the simple case of four emitters, we require six free parameters but have only the energy and three detunings. As such, the addition of two external couplings allows to stabilize any two-excitation Lindbladian dark state, as shown in Appendix~\ref{Appendix: 4atoms_2exc}. While this method allows for perfect stabilization, the required number of couplings scales quickly with the number of excitations of the dark state. External interactions can also solve the problem of divergent optimal detunings. For the case of two emitters, for example, this is achieved by applying an external coupling that exactly cancels the waveguide-mediated interactions. In that case, a perfect dark state exists even for $x_2 = \lambda_0 n /2$ without the need of applying local detunings. 

\section{Deterministic generation of dark states}
\label{section: dark_state_generation}

The dark states stabilized in Section~\ref{section: dark_state_engineering} can be prepared in multiple ways.  The efficient preparation of delocalized dark states can be achieved by weak classical driving fields external to the waveguide (\ie through sideports in superconducting qubits or through free space for atoms trapped close to nanophotonic structures), which can have a different amplitude and phase profile than the waveguide field and can therefore directly couple to dark states. As we show in Section~\ref{section: dark_state_generation_classical}, this results in high efficiency Rabi oscillations between dark states provided that the remaining states in adjacent excitation manifolds are bright (Zeno effect) or far off-resonant (photon blockade)~\cite{Rubies25}. Alternatively, dark states can also be prepared by storing few-photon pulses via a controlled coupling to bright states, as demonstrated in Section~\ref{section: dark_state_generation_photons}. Finally, localized dark states can be produced probabilistically via heralding with high-efficiency~\cite{Holzinger22}, as presented in Appendix~\ref{Appendix: heralded_dissipation}. 

\subsection{Classical driving fields}
\label{section: dark_state_generation_classical}

\begin{figure*}   
\includegraphics[width=\linewidth]{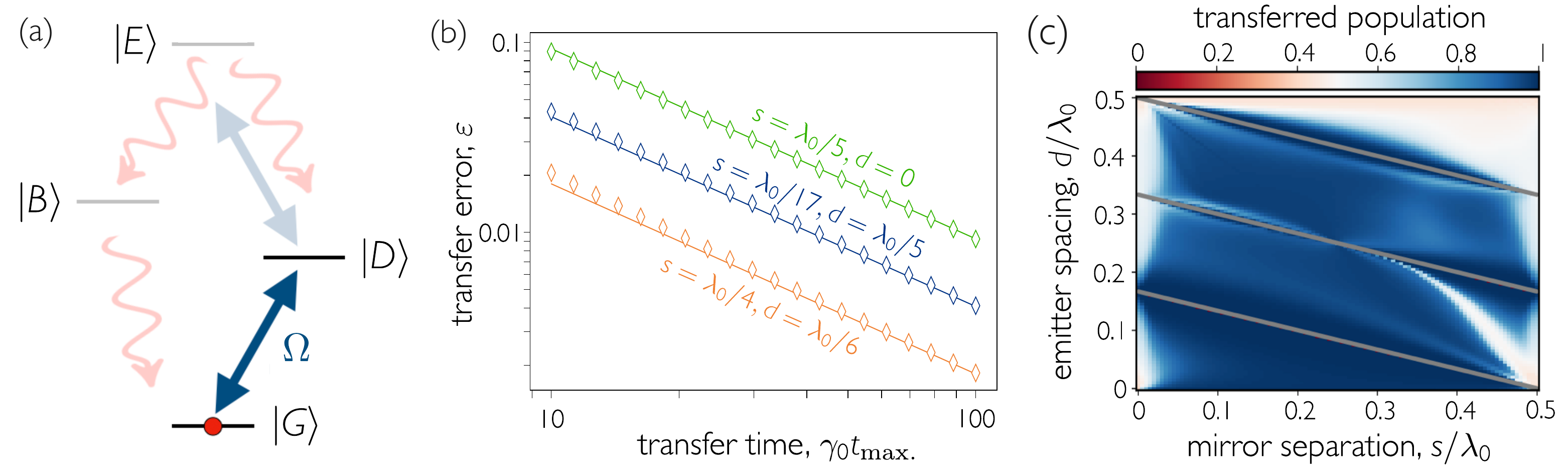}
    \caption{(a) Collective states of a two-emitter system at the optimal detuning given in Eq.~(\ref{eq: optimal_detuning_2atoms}). For weak resonant drive on the $|G\rangle \leftrightarrow |D\rangle$ transition, the bright states can be adiabatically eliminated due to the Zeno effect and the dipole-induced photon blockade, leading to high fidelity Rabi oscillations between $|G\rangle $ and $|D\rangle$. (b) Error of the dark state preparation protocol, $\epsilon = 1-p_D^{(max)}$, as a function of the transfer time $t_\mathrm{max}$ for two emitters with various configurations $s$ and $d$. The diamonds and solid lines respectively correspond to the solution of the full master equation and the adiabatic analytical result in Eq.~(\ref{eq: scaling_error}). A scaling $\epsilon \sim t_\mathrm{max}^{-1}$ is observed. (c) Maximum transferred population to the target single-excitation dark state, $ |\psi_\mathcal{D}^{(1)} \rangle \propto |e_1\rangle + |e_2\rangle + |e_3\rangle - [\sin (k_0x_1) + \sin (k_0x_2) + \sin (k_0x_3) ] |e_4\rangle / \sin (k_0x_4)$, for a system with four emitters as a function of $s$ and $d$ for a driving strength $\Omega = \pi \gamma_0/50$ and the system initially in the ground state. }
    \label{fig: generate_classical_drive}
\end{figure*}

For a system of two emitters, the eigenstates of the non-Hermitian Hamiltonian under the optimal detuning in Eq.~(\ref{eq: optimal_detuning_2atoms}) are the ground state $|G \rangle$ and the single-excitation dark state $|D\rangle$ [Eq.~(\ref{eq: 2atoms_darkstate})] which do not decay, and the fully excited state $|E \rangle$ and the bright state $|B\rangle$ [Eq.~(\ref{eq: 2atoms_brightstate})] which both decay at a rate $\Gamma_B$. An applied external drive adds the Hamiltonian term
\begin{equation}
\label{eq: drive_dark_1exc_2at}
\hat{H}_\mathrm{d} = \Omega 
\frac{\sin (k_0 x_2) \hat{\sigma}_1^{eg} - \sin (k_0 x_1) \hat{\sigma}_2^{eg}}{\sqrt{\sin (k_0 x_1)^2+\sin (k_0 x_2)^2}} e^{-i \omega_L t}  + h.c.,
\end{equation} 
which couples the  $|G\rangle \leftrightarrow |D\rangle$ transition with strength $\Omega$, as well as the $|D\rangle \leftrightarrow |E\rangle$ and $|B\rangle \leftrightarrow |E\rangle$ transitions.

We consider the system initially in $|G\rangle$ and the drive resonant with the $|G\rangle \leftrightarrow |D\rangle$ transition. Then, the drive is off-resonant with the $|D\rangle \leftrightarrow |E\rangle$ transition by $\Delta_E - 2\Delta_D$, where $\Delta_\lambda$ is the detuning of state $|\lambda \rangle$ from the bare emitter resonance and is given in Appendix~\ref{Appendix: Generating_dark_classical_drive}. For a weak driving field, two distinct processes suppress the population of $|E\rangle$: the large decay rate of $|E\rangle$, leading to a Zeno effect~\cite{Facchi02} for $\Omega \ll \Gamma_B$; and the off-resonant drive, leading to a dipole-induced photon blockade~\cite{Cidrim20,Williamson20PRL} for $|\Delta_E - 2\Delta_D| \gg \Omega$. In this regime, we can adiabatically eliminate the bright states $|E\rangle$ and $|B\rangle$, as illustrated in Fig.~\ref{fig: generate_classical_drive}(a), and obtain the effective evolution for the dark subspace of the system, which exhibits resonant Rabi oscillations between $|G\rangle$ and $|D\rangle$. The bright states additionally induce radiative decay from $|D\rangle$ to $|G\rangle$ and pure dephasing of the state $|D\rangle$, which result in non-unit population transfer to the single-excitation dark state. As shown in Appendix~\ref{Appendix: Generating_dark_classical_drive}, the error of the dark state preparation protocol,
\begin{equation}
\label{eq: scaling_error}
    \epsilon \approx \frac{\pi^2}{32}   \frac{ b^2(1\!+\!2 b^2) \Gamma_B }{(\Delta_E \! -\!2\Delta_D)^2 + (\Gamma_B/2)^2} t_\mathrm{max}^{-1} + \mathcal{O} (t_\mathrm{max}^{-2}),
\end{equation}
scales inversely proportional to the duration of the light pulse, $t_\mathrm{max} \approx \pi / 2 \Omega$. Here, we have defined $b=2 \sin (k_0x_1)\sin (k_0x_2)/[\sin^2 (k_0x_1)+\sin^2 (k_0x_2)]$. Equation~(\ref{eq: scaling_error}) is in good agreement with a full solution of the master equation, as shown in Fig.~\ref{fig: generate_classical_drive}(b) for various configurations. Notably, the error diminishes for configurations where the Zeno effect and the photon-blockade become more prominent (\ie when $\Gamma_B$ and $|\Delta_E-2\Delta_D|$ increase). In fact, small errors are obtained for $t_\mathrm{max} \sim 20 / \gamma_0$ for all configurations except for those where (i) $\Gamma_B\to 0$ and $\Delta_D=0$ such that the required timescale for adiabaticity diverges (\ie for $s$ and $d$ close to $m\lambda_0/2$ with $m \in \mathbb{N}$), and where (ii) the single-excitation state cannot be stabilized due to divergent optimal detunings (\ie for $x_2 = s + d \approx \lambda_0 n /2 $ with $n \in \mathbb{N}$).

The two-emitter case is readily generalized to $N$ emitters. As before, the target total dark state $|\psi_\mathcal{D}^{(1)}\rangle = \sum_{j=1}^N \alpha_j  |e_j\rangle$ given by Eq.~(\ref{eq: 1exc_dark_Natoms}) and stabilized by the optimal detunings in Eq.~(\ref{eq: optimaldet_Natoms_1exc}) is generated by a weak drive that matches its structure,
\begin{equation}
\hat{H}_\mathrm{d} = \Omega \sum_j \alpha_j \hat{\sigma}_j^{eg} + h.c. 
\end{equation}
To attain approximately closed dynamics on the $|G \rangle \leftrightarrow  |\psi_\mathcal{D}^{(1)} \rangle$ transition, we require that all the two-excitation states coupled to $ |\psi_\mathcal{D}^{(1)} \rangle$ by the dark drive are either bright or out of resonance. As occurred for the case of two emitters, excellent transfer with error scaling as $\epsilon \propto t_\mathrm{max}^{-1}$ (see Appendix~\ref{Appendix: Generating_dark_classical_drive}) is achieved for most configurations except when the optimal detuning diverges and when there exists a resonant transition to two-excitation dark states [Fig.~\ref{fig: generate_classical_drive}(c)].

Separable multi-excitation states can be prepared in a simple manner. For an $n_\mathrm{exc}$-excitation dark state of the form $|\psi_\mathcal{D}^{(n_\mathrm{exc})} \rangle = \bigotimes_{b=1}^{n_\mathrm{exc}} |\psi_b \rangle$, it suffices to drive the single-excitation states $|\psi_b \rangle$ of the $n_\mathrm{exc}$ subsystems sequentially and at their respective resonance frequencies. An example of the preparation of the four-emitter dark state $| D_{12,34} \rangle =| D_{12} \rangle \otimes | D_{34} \rangle $ given in Eq.~(\ref{eq: darkstate_2exc_separable}) with an efficiency of $96 \%$ is given in Appendix~\ref{Appendix: Generating_dark_classical_drive}.

More general optimized multi-excitation dark states are harder to prepare with classical driving fields. While a dark driving field can couple the ground state of the system $|G\rangle$ to just one single-excitation dark state $| \psi_\mathcal{D}^{(1)} \rangle$, it cannot cleanly drive (for general configurations) the transition $| \psi_\mathcal{D}^{(1)} \rangle \leftrightarrow | \psi_\mathcal{D}^{(2)} \rangle$ without directly populating other states within the manifolds with one and two excitations. As a result, the likelihood to directly address dark resonant states increases and the efficiency of the preparation scheme is reduced.

\subsection{Via few-photon absorption}
\label{section: dark_state_generation_photons}

\begin{figure}
    \includegraphics[width=\columnwidth]{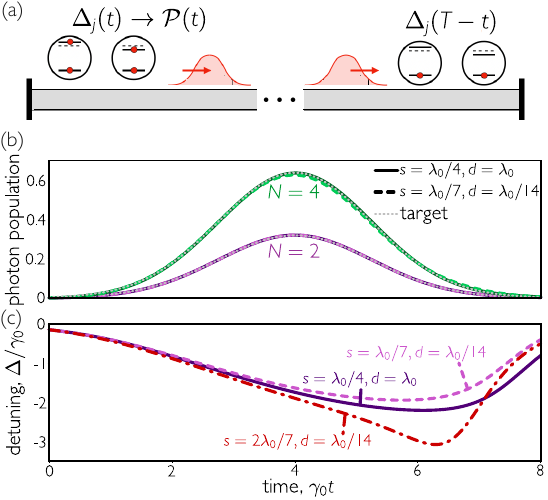}
    \caption{(a) Retrieval and absorption of a symmetric Gaussian photon $\mathcal{P}(t)$ with duration $T$ by applying the optimized detuning sequence $\Delta_j(t)$ and $\Delta_j(T-t)$, respectively. (b) Shape $\mathcal{P}(t)$ of the retrieved single- and two-photon pulses for two different configurations, respectively stored in the (purple) two-emitter dark state $|D\rangle$ and (green) the four-emitter product dark state $|D_{12} \rangle \otimes |D_{34} \rangle$. The gray dashed lines correspond to the target Gaussian photons, $\mathcal{P}_G(t)$ with duration $T=8 / \gamma_0$ and width $\tau = 1.25 /\gamma_0$. (c) Time-dependent detuning resulting in a Gaussian photon for $N=2$ in different setups. For the four-emitter system with $s=\lambda_0/4$ and $d=\lambda_0$, the time-dependent detuning between the first two and the last two emitters is given by the dark purple line. For the four-emitter system with $s=\lambda_0/7$ and $d=\lambda_0/14$, the time-dependent detunings between the first two and the last two emitters are given by the light purple and red curves, respectively. We consider a detuning between both subsystems of $2\Delta_\mathrm{off} = 10 \gamma_0$. 
}
    \label{fig: memory}
\end{figure}

Dark states with $n_\mathrm{exc}$ excitations can also be prepared by absorbing a light pulse with $n_\mathrm{exc}$ photons from the half-waveguide. Since dark states are decoupled from the electromagnetic field of the half-waveguide, coupling to radiating or bright collective states is required for the absorption protocol. In particular, we controllably destabilize the dark state by altering the resonance frequency or detuning $\Delta_j(t)$ of each individual emitter over time. Photon absorption and retrieval are time-reversed processes of each other: if applying $\Delta_j(t)$ results in the emission of a photon with temporal shape $\mathcal{P}(t)$ (\ie the photonic population emitted into the waveguide per unit of time) and duration $T$, then the time-reversed photon $\mathcal{P}(T-t)$ can be perfectly absorbed by applying the time-reversed detuning $\Delta_j(T-t)$ [see Fig.~\ref{fig: memory}(a)]. We can thus recast the dark state preparation problem as an emission problem where we aim to find the detuning sequence $\Delta_j(t)$ that results in a photon with temporal shape $\mathcal{P}(t)$.

We return to the case of two emitters as an illustrative example, and consider an excitation initially in the total dark state in Eq.~(\ref{eq: 2atoms_darkstate}) under the optimal relative detuning $\Delta_\mathrm{opt}$ in Eq.~(\ref{eq: optimal_detuning_2atoms}). The photonic population emitted into the half-waveguide per unit of time, $\mathcal{P} (t) = \Gamma_B |e_B(t)|^2 $, determines the required population in the bright state $|B\rangle$ over time~\cite{Rubies22PRR,Rubies25}. This fixes the coupling strength between $|D\rangle$ and $|B\rangle$, which in turn determines the relative detuning $\Delta_1(t) - \Delta_2(t) = \Delta_\mathrm{opt} + \Delta(t) $ that needs to be applied over time. The resulting analytical form for the temporal detuning profile $\Delta(t)$ is given in Appendix~\ref{Appendix: few_photon_absorption}. As an example, we show that a desired Gaussian photon of duration $T$ and width $\tau$, $\mathcal{P}_G(t) = \exp \left( -(t-T/2)^2/ 2 \tau^2 \right) / \sqrt{\tau \pi}$, can be obtained to high fidelity in Fig.~\ref{fig: memory}(b) using the detuning profile shown in Fig.~\ref{fig: memory}(c). Note that the timescale of the emitted photon is limited by the inverse decay rate $\Gamma_B^{-1}$ of the bright jump operator. 

Similarly, we can store and retrieve $n_\mathrm{exc}$ photons using the $n_\mathrm{exc}$ dark product states. As an example, two photons can be stored in the four-emitter dark state $|D_{12,34}\rangle =|D_{12} \rangle \otimes |D_{34} \rangle$ introduced in Sec.~\ref{subsection: engineeringdark_multi}. For this dark state to be stabilized, the two subsystems are significantly detuned from each other by $2\Delta_\mathrm{off}$, and one excitation is stored in each subsystem. As such, emission of photons follows the same protocol as above for each subsystem. We show that this can be used to produce Gaussian pulses with precisely two photons in Fig.~\ref{fig: memory}(b). Of course, the two photons can also be emitted in distinct temporal profiles.

For single-excitation dark states with more than three emitters or multi-excitation dark states that cannot be written as product states of single-excitation dark states, the optimal detuning sequence can in general be found through numerical optimization. As we show in Appendix~\ref{Appendix: few_photon_absorption}, the storage and emission protocols are robust to noise in the applied temporal detuning sequence, making dark states reliable high-fidelity photon memories and quantum network nodes.

\section{Conclusions and Outlook}

In this paper, we have studied arrays of emitters coupled to a half-waveguide. While the dissipative coupling of each emitter to the waveguide depends on its exact position, photon emission is governed by a single collective jump operator. The waveguide field additionally mediates coherent interactions between the emitters, which prevent the existence of perfect dark states in arbitrary configurations due to mixing with radiating states. We further demonstrate that the effect of coherent interactions can be canceled by applying optimized local frequency shifts, which stabilize perfect single-excitation and near-perfect multi-excitation dark states. We further show that these dark states can be deterministically prepared with a weak classical driving field external to the waveguide, and by absorbing few-photon pulses that propagate through the waveguide. 

Our results could be readily implemented with superconducting qubits coupled to a common coplanar waveguide~\cite{Hoi15,Zanner22,Mirhosseini19,Kannan23,Ferreira24,OSullivan25,Du26arxiv}, as well as with neutral atoms or quantum dots coupled via nanophotonic waveguides~\cite{Goban14,Goban15,Tiranov23,Ocola24}. Individual addressing of each emitter (\ie control over its resonance frequency and driving strength at each position) can be achieved through separate control lines in the former case, and free space driving fields in the latter.

Selective coupling to radiating states via local frequency shifts enables the storage of few-photon pulses with arbitrary wavepackets in multi-excitation dark states, as well as their controlled release. Beyond serving as quantum memories for light~\cite{Asenjo17PRX,Manzoni18,Rubies22PRR,Ballantine21}, this capability transforms emitter arrays into versatile nodes for quantum networking and communication protocols~\cite{Cirac97,Duan01,Knaut24,Almanakly25}, and for the generation of entangled states of light~\cite{Rubies25}. This work can be directly extended to other platforms, such as arrays of emitters coupled to a (full) one-dimensional waveguide~\cite{Lalumiere13,Holzinger22}. In that case, the system generally exhibits two jump operators and would therefore require three emitters to attain a perfect dark state of the full master equation upon applying the optimized frequency shifts. Finally, it is worth noting that the lifetime of the dark and subradiant states discussed here will be limited by decoherence and undesired free-space decay in realistic experimental implementations. For closely spaced emitters, the free-space vacuum field induces additional coherent and dissipative interactions. This opens the possibility of engineering collective states that are dark with respect to the waveguide and subradiant with respect to free space, thereby reducing undesired losses~\cite{Asenjo17PRX}.

\begin{acknowledgments}
We would like to thank Manon Revel, Ricardo Guti\'{e}rrez-J\'{a}uregui, Stefan Ostermann, Valentin Walther and  Yidan Wang for fruitful discussions.
O.R.-B. acknowledges support from Fundación Mauricio y Carlota Botton. A. A.-G. acknowledges support by the National Science Foundation through the CAREER Award (No. 2047380),  the U.S. Department of Energy, Office of Science, National Quantum Information Science Research Centers, Co-design Center for Quantum Advantage (C2QA) under Contract No.~DE-SC0012704, as well as by the Chu Family Foundation and the David and Lucile Packard Foundation. SFY acknowledges support by the NSF through the CUA PFC  (PHY-2317134) and PHY-2207972 in addition to AFOSR through FA9550-24-1-0311.

\end{acknowledgments}

\vspace{0.5cm}
\appendix

\section{Number of Lindbladian dark states}
\label{appendix: numberDarkStates}

Here, we derive bounds regarding the number of Lindbladian dark states $|\psi_\mathcal{D}^{n_\mathrm{exc}}\rangle$ for different excitation manifolds. These states do not decay under the action of the jump operator $\hat{S}_B$, and therefore fulfill $\hat{S}_B |\psi_\mathcal{D}^{n_\mathrm{exc}}\rangle  = 0$ or equivalently $\hat{H}_\mathrm{dis} |\psi_\mathcal{D}^{n_\mathrm{exc}}\rangle = 0$. \\

\emph{\underline{Proposition A1}: There are at least $ \max \{ 0, {N\choose n_\mathrm{exc}} - {N\choose n_\mathrm{exc}-1} \}$ Lindbladian dark states (\ie eigenstates of the dissipative interactions $\hat{H}_\mathrm{dis} = \Gamma_B \hat{S}_B^\dagger \hat{S}_B / 2$ with eigenvalue zero) with $n_\mathrm{exc}$ excitations.}

\underline{Proof:} A general state with $n_\mathrm{exc}$ excitations can be expressed as a superposition of ${N\choose n_\mathrm{exc}}$ basis states, $| \psi \rangle = \sum_{j_1<j_2<\dots<j_{n_\mathrm{exc}}} \alpha_{j_1,j_2,\dots,j_{n_\mathrm{exc}}} |e_{j_1} e_{j_2} \dots e_{j_{n_\mathrm{exc}}} \rangle$, where $\alpha_{j_1,j_2,\dots,j_{n_\mathrm{exc}}}$ are ${N\choose n_\mathrm{exc}}$ free parameters. The dimension of the dark subspace with $n_\mathrm{exc}$ is equal to the number of states that fulfill $\hat{S}_B |\psi_\mathcal{D}^{n_\mathrm{exc}}\rangle = 0$. This condition results in a set of ${N\choose n_\mathrm{exc}-1}$ equations and ${N\choose n_\mathrm{exc}}$ parameters or unknowns. The dimension of the dark subspace is equal to the number of unknowns minus the number of linearly independent equations, and is minimal if all equations are linearly independent. Thus, the minimum number of Lindbladian dark states is ${N\choose n_\mathrm{exc}} - {N\choose n_\mathrm{exc}-1}$ if $n_\mathrm{exc} \leq N/2$ and 0 otherwise. \\

\emph{\underline{Proposition A2}: If all emitters are bright, there are exactly $ \max \{ 0, {N\choose n_\mathrm{exc}} - {N\choose n_\mathrm{exc}-1} \}$ Lindbladian dark states with $n_\mathrm{exc}$ excitations for  $n_\mathrm{exc}=1$ and $n_\mathrm{exc}=2$.}

\underline{Proof:} Let us first consider the single-excitation manifold, $n_\mathrm{exc}=1$. Acting $\hat{S}_B^\dagger$ on the zero-excitation manifold (\ie the ground state $|G\rangle$) results in a single state, $|B \rangle \propto \hat{S}_B^\dagger |G \rangle \propto \sum_j \sin(k_0 x_j) |e_j\rangle$. The orthogonal subspace has dimension $N\!-\!1$, $\mathcal{D}_1 \equiv \{ |\psi_m \rangle \}$, where $m \in [1,N\!-\!1]$. If at least one emitter is bright (\ie $\sin(k_0 x_j)\neq 0$), then $\langle B| \hat{S}_B^\dagger |G \rangle= \langle B | B \rangle \neq 0$. Then, $\langle G| \hat{S}_B |B \rangle = ( \langle B| \hat{S}_B^\dagger |G \rangle )^\dagger \neq 0$ and $|B\rangle$ decays to the ground state by emitting photons. The orthogonal subspace, however, fulfills $0 = \langle \psi_m | B \rangle = \langle \psi_m | \hat{S}_B^\dagger |G \rangle$. Applying the Hermitian conjugate again, it follows that $\langle G | \hat{S}_B | \psi_m\rangle = 0$, and the $N\!-\!1$ states $\{ |\psi_m \rangle \}$ in subspace $\mathcal{D}_1$ are dark and do not decay. 

A similar proof holds for the two-excitation subspace with $N\geq3$. Given the single-excitation basis $\{ |e_j\rangle \}$ with $j \in [1,N]$, we can build the two-excitation subspace spanned by $\hat{S}_B^\dagger$ acting on $\{ |e_j\rangle \}$, $\mathcal{B}_2 = \{ |\phi_j \rangle = \hat{S}_B^\dagger |e_j \rangle \propto \sum_{l \neq j} \sin(k_0 x_l) |e_j e_l \rangle \}$. If all emitters are bright ($\sin(k_0 x_l) \neq 0 ~\forall l$), $\mathcal{B}_2$ has dimension $N$. For that, we demonstrate that all states $|\phi_j \rangle$ are linearly independent and that the only scalar combination $\alpha_j$ such that $\sum_j \alpha_j |\phi_j \rangle = 0$ is $\alpha_j =0$ for all $j$. For example, the amplitudes on the states $|e_1 e_j \rangle$ need to be canceled for all $j \neq 1$. Their only contribution come from states $|\phi_1 \rangle$ and $|\phi_j \rangle$, and result in the set of equations $0=\alpha_1 \sin(k_0 x_j) + \alpha_j \sin(k_0 x_1)$. Thus, $\alpha_j = - \alpha_1 \sin(k_0 x_j) /\sin(k_0 x_1)$. Additional cancellation of the amplitude in the state $|e_2 e_3\rangle$ results in the equation $0 = \alpha_2 \sin(k_0 x_3) + \alpha_3 \sin(k_0 x_2) = -2 \alpha_1 \sin(k_0 x_2) \sin(k_0 x_3) /\sin(k_0 x_1)$, which trivially results in $\alpha_1 =0 $ and consequently $\alpha_j =0$ for all $j$, and the proof is done. 

Next, we need to show that the states in subspace $\mathcal{B}_2$ are bright and decay. For that, we note that $\langle \phi_j | \phi_j \rangle = \langle \phi_j | \hat{S}_B^\dagger |e_j \rangle \neq 0 $ if $\sin (k_0 x_j) \neq 0$ for all $j$. Applying Hermitian conjugation, we find $ \langle e_j |\hat{S}_B | \phi_j \rangle \neq 0$, and states in $\mathcal{B}_2$ decay.

Finally, we can construct the subspace perpendicular to $\mathcal{B}_2$, $\mathcal{D}_2 = \{ | \psi_j \rangle \}$. Since there are $N(N-1)/2$ two-excitation states, $\mathcal{D}_2$ has dimension $N(N-3)/2$. Using $\langle \phi_j | \psi_l \rangle = 0$ for all $j$ and $l$, and applying the definition of $| \phi_j \rangle$, it follows that $\langle e_j | \hat{S}_B | \psi_l \rangle =0$ for all $j$ and $l$. Thus, all $N(N-3)/2 = {N\choose 2} - {N\choose 1} $ states in $\mathcal{D}_2$ do not decay and are dark. \\

\emph{\underline{Proposition A3}: There only exist Lindbladian dark states with $N-1$ excitations if at most 2 emitters are bright.}

\underline{Proof:} A general state with $N-1$ excitations can be written as $| \psi \rangle = \sum_j \alpha_j |g_j\rangle$, where $|g_j\rangle$ denotes the state where only emitter $j$ is in the ground state. $| \psi \rangle$ is a dark state of $H_\mathrm{dis}=\Gamma_B \hat{S}_B^\dagger \hat{S}_B /2$ only if $\hat{S}_B | \psi \rangle \propto \sum_{j < l} (\alpha_j \sin(k_0 x_l) + \alpha_l \sin(k_0 x_j)) |g_j g_l\rangle =0$, that is, if
\begin{equation}
\label{eq: appendix_number_N-1}
    \alpha_j \sin(k_0 x_l) + \alpha_l \sin(k_0 x_j)=0
\end{equation}
$\forall ~j \neq l \in [1,N]$. Let us consider the following scenarios:

(a) All emitters are dark ($\sin(k_0 x_j) = 0$ $\forall~ j$). Then, Eq.~(\ref{eq: appendix_number_N-1}) is trivially fulfilled $\forall j,l \in [1,N]$ independently of $\alpha_j$, and all states in the $N-1$ excitation manifold are dark.

(b) Only one emitter is bright ($\sin(k_0 x_1) \neq 0$ and $\sin(k_0 x_j)=0$ $\forall j \neq 1$). Then, Eq.~(\ref{eq: appendix_number_N-1}) reduces to $\alpha_l=0$ $\forall l \neq 1$, and there exists one dark states with $N-1$ excitations, $|\psi \rangle = |g_1\rangle$.

(c) Only two emitters are bright ($\sin(k_0 x_j) \neq 0$ only for $j=1$ and $j=2$). Then, Eq.~(\ref{eq: appendix_number_N-1}) reduces to $\alpha_l=0$ $\forall l \neq 1,2$ and $\alpha_1 \sin(k_0 x_2) + \alpha_2 \sin(k_0 x_1)=0$, and there exists one dark state, $|\psi \rangle \propto \sin(k_0 x_1) |g_1\rangle -\sin(k_0 x_2) |g_2\rangle = (\sin(k_0 x_1) |ge\rangle -\sin(k_0 x_2) |eg\rangle) \otimes |e\rangle^{\otimes (N-2)} $.

(d) More than two emitters are bright. If emitters $1$ and $2$ are bright, we can write $\alpha_j = - \alpha_1 \sin(k_0 x_j)/\sin(k_0 x_1) = -\alpha_2 \sin(k_0 x_j)/\sin(k_0 x_2)$ [see Eq.~(\ref{eq: appendix_number_N-1})]. Since $\sin(k_0 x_j) \neq 0$ for at least one $j$, this results in $\alpha_1 \sin(k_0 x_2) - \alpha_2 \sin(k_0 x_1)=0$. Equation~(\ref{eq: appendix_number_N-1}) also results in the condition $\alpha_1 \sin(k_0 x_2) + \alpha_2 \sin(k_0 x_1)=0$. For $\sin(k_0 x_1) \neq 0$ and $\sin(k_0 x_2) \neq 0$, the only solution is $\alpha_1=\alpha_2=0$. Then, $\alpha_j=0$ $\forall j$, and there does not exist any dark state.

In conclusion, a dark state of $H_\mathrm{dis}$ with $N-1$ excitations only exists if at least $N-2$ emitters are dark, such that the problem reduces to finding a single-excitation Lindbladian dark states with two emitters only. \\

\emph{\underline{Proposition A4}: There only exist Lindbladian dark states with $N-2$ excitations if at most 4 emitters are bright.}

\underline{Proof:} A general state with $N-2$ excitations, $| \psi \rangle = \sum_{j<l} \alpha_{jl} |g_j g_l\rangle$, is a dark state of $H_\mathrm{dis}=\Gamma_B \hat{S}_B^\dagger \hat{S}_B /2$ only if $\hat{S}_B | \psi \rangle = 0$, which results in the condition
\begin{equation}
\label{eq: appendix_number_N-2}
    \alpha_{jl} \sin(k_0 x_m) + \alpha_{lm} \sin(k_0 x_j) + \alpha_{jm} \sin(k_0 x_l) =0
\end{equation}
$\forall ~j\neq l \neq m \in [1,N]$. Let us consider the following scenarios:

(a) All emitters are dark ($\sin(k_0 x_j)=0$ $\forall j$). Then, Eq.~(\ref{eq: appendix_number_N-2}) is trivially fulfilled and all ${ N \choose 2}$ states in the manifold are dark.

(b) Only one emitter is bright ($\sin(k_0 x_1) \neq 0$ and $\sin(k_0 x_j)=0$ $\forall~ j \neq 1$). Then, Eq.~(\ref{eq: appendix_number_N-2}) reduces to $\alpha_{jl}=0$ $\forall j,l \neq 1$, and there exist $N-1$ dark states of the form $|\psi \rangle = \sum_{j \neq 1} \alpha_{1j} | g_1 g_j \rangle$.

(c) Only two emitters are bright ($\sin(k_0 x_j) \neq 0$ only for $j=1$ and $j=2$). Taking $j=1$ and $l,m>2$ in Eq.~(\ref{eq: appendix_number_N-2}), we find $\alpha_{lm} = 0$ $\forall~ l,m > 2$. Taking now the term $j=1$ and $l=2$, it follows $\alpha_{1m} \sin(k_0 x_2) + \alpha_{2m} \sin(k_0 x_1) = 0$ $\forall ~m$. Thus, there exist $N-1$ Lindbladian dark states in total. Among them, $N-2$ are of the form $| \psi \rangle = ( |g_1 \rangle - |g_2 \rangle \sin(k_0 x_2)/\sin(k_0 x_1))  \otimes \sum_{m =3}^N \alpha_{1m} |g_m \rangle$, each corresponding to the single-excitation dark state involving the two bright emitters, with one of the remaining dark emitters in its ground state. The remaining Lindbladian dark state is $|g_1g_2\rangle$, in which both bright emitters are in their ground states.

(d) Only three emitters are bright ($\sin(k_0 x_j) \neq 0$ only for $j<4$). Following a similar procedure as above, we find that the only non-zero weights are $\alpha_{12}$, $\alpha_{13}$ and $\alpha_{23}$, and that Eq.~(\ref{eq: appendix_number_N-2}) is only fulfilled for two states, $| \psi \rangle = \alpha_{12} |g_1 g_2 \rangle + \alpha_{13} |g_1 g_3 \rangle - (\alpha_{12} \sin(k_0 x_3)/\sin(k_0 x_1) + \alpha_{13} \sin(k_0 x_2)/\sin(k_0 x_1)) |g_2 g_3 \rangle$.

(e) Only four emitters are bright ($\sin(k_0 x_j) \neq 0$ only for $j<5$). Now the only nonzero amplitudes are $\alpha_{jl}$ such that $j \neq l<5$. Equation~(\ref{eq: appendix_number_N-2}) results in a set of 6 unknowns and 4 equations, identical to the one solved for two excitations and four bright emitters in Section~\ref{subsection: engineeringdark_multi}. Its solution are two Lindbladian dark states with free parameters $\alpha_{24}$ and $\alpha_{34}$, and all other amplitudes
    \begin{subequations}
\begin{align}
        \alpha_{12} &= \frac{\alpha_{34} \sin(k_0 x_1) \sin(k_0 x_2)}{\sin(k_0 x_3) \sin(k_0 x_4)} \\
        \alpha_{13} &= \frac{\alpha_{24} \sin(k_0 x_1) \sin(k_0 x_3)}{\sin(k_0 x_2) \sin(k_0 x_4)} \\
        \alpha_{23} &= \frac{-\alpha_{24} \sin(k_0 x_3) -\alpha_{34} \sin(k_0 x_2)}{ \sin(k_0 x_4)} \\ 
        \alpha_{14} &=  -\sin(k_0 x_1)\left[ \frac{\alpha_{34}}{\sin(k_0 x_3)} +\frac{\alpha_{24}} { \sin(k_0 x_2)} \right]
\end{align}
\end{subequations}.

(f) Only five emitters are bright ($\sin(k_0 x_j) \neq 0$ only for $j<6$). Now, Eq.~(\ref{eq: appendix_number_N-2}) results in a set of 10 nonzero amplitudes or unknowns, $\alpha_{jl}$ such that $j \neq l<6$, and 10 equations (corresponding to all permutations of $j \neq l \neq m <6$). The linear system can be written in terms of a matrix with determinant $-48 \prod\limits_{j=1}^5 \sin^2(k_0 x_j)\neq 0$, which is therefore invertible. Thus, the only solution is the trivial one, $\alpha_{jl} =0$ $ \forall j \neq l<6$, and there exists no Lindbladian dark state.

(g) Six or more emitters are bright. The non-zero amplitudes can only be $\alpha_{jl}$ such that $j\neq l$ are both bright. Taking a combination of $5$ bright emitters, we have readily shown that their set of equations result in their $10$ amplitudes being zero. That is true for any other combination of five bright emitters, which trivially results in all amplitudes being zero. Thus, no Lindbladian dark states are possible.

In conclusion, a dark state of $H_\mathrm{dis}$ with $N-2$ excitations only exists if at least $N-4$ emitters are dark.

\section{Dissipation from Lindbladian eigenstates}
\label{appendix: proof_jump}

In this Appendix, we study the nature of the jump operator $\hat{S}_B$ and elucidate its action on eigenstates of the dissipative interactions. \\ 

\emph{\underline{Proposition B1}: The action of the jump operator $\hat{S}_B$ on an eigenstate of the dissipative interactions $\hat{H}_\mathrm{dis} = \Gamma_B \hat{S}_B^\dagger \hat{S}_B/2$ with $n_\mathrm{exc}$ excitations necessarily results in an eigenstate of $\hat{H}_\mathrm{dis}$ with $n_\mathrm{exc}-1$ excitations only for the generalized Dicke model ($d=m\lambda_0/2$ with $m \in \mathbb{N}$).}

\underline{Proof}: Let us consider the jump operator $\hat{S}_B$ given by Eq.~(\ref{eq: jump_operator}). To simplify notation, we define $\hat{S}= \sqrt{\Gamma_B/2\gamma_0} \hat{S}_B = \sum_j \sin(k_0 x_j) \hat{\sigma}_j^{ge}$. From the spin-nature of the two-level emitters, which fulfill $ [\hat{\sigma}_j^{eg},\hat{\sigma}_j^{ge}] = \hat{\sigma}_j^z$ with $\hat{\sigma}_j^z = \hat{\sigma}_j^{ee}-\hat{\sigma}_j^{gg} = |e_j \rangle \langle e_j| -|g_j \rangle \langle g_j|$, it follows that 
\begin{equation}
\label{eq: app_commutation_S}
    \hat{S} \hat{S}^\dagger = \hat{S}^\dagger \hat{S} - \sum_j \sin^2(k_0 x_j) \hat{\sigma}_j^z.
\end{equation}

Let us additionally consider an eigenstate $|\alpha \rangle$ of the dissipative interactions with $n_\mathrm{exc}$ excitations, such that $\hat{S}^\dagger \hat{S} |\alpha \rangle= \alpha |\alpha \rangle $ and $\sum_j \hat{\sigma}_j^{ee} |\alpha \rangle = n_\mathrm{exc} |\alpha \rangle$. Acting the jump operator on $|\alpha \rangle$, we obtain $|\beta \rangle = \hat{S} |\alpha \rangle$. $|\beta \rangle$ is also an eigenstate of the dissipative interactions if and only if $\hat{S}^\dagger \hat{S} |\beta \rangle= \beta |\beta \rangle $ with $\beta \in \mathbb{R}$. Using Eq.~(\ref{eq: app_commutation_S}), we can write
\begin{align}
   \hat{S}^\dagger \hat{S} |\beta \rangle &=  (\hat{S} \hat{S}^\dagger + \sum_j \sin^2(k_0 x_j) \hat{\sigma}_j^z ) \hat{S} |\alpha \rangle \\&= \alpha \hat{S} |\alpha \rangle + \sum_j \sin^2(k_0 x_j) \hat{\sigma}_j^z \hat{S} |\alpha \rangle,
\end{align}
and the problem reduces to finding whether $|\alpha \rangle$ fulfills 
\begin{align}
\label{eq: app_jump_condition}
   \sum_j \sin^2(k_0 x_j) \hat{\sigma}_j^z \hat{S} |\alpha \rangle = \tilde{\beta} |\beta \rangle =\tilde{\beta} \hat{S} |\alpha \rangle.
\end{align} 
For the generalized Dicke model, $\sin^2(k_0 x_j)$ is equal for all emitters $j$ (\ie $\sin^2(k_0 x_j) = \sin^2(k_0 x_1)$). Additionally considering that $\hat{S} | \alpha \rangle$ is a state with $n_\mathrm{exc}-1$ excitations, $\sum_j \sin^2(k_0 x_j) \hat{\sigma}_j^z \hat{S} |\alpha \rangle = \sin^2(k_0 x_1) (2n_\mathrm{exc}-2-N) \hat{S} |\alpha \rangle$ and Eq.~(\ref{eq: app_jump_condition}) is fulfilled. That is, applying the jump operator on an eigenstate of $\hat{H}_\mathrm{dis}$ necessarily results in another eigenstate of $\hat{H}_\mathrm{dis}$ if the system reduces to the generalized Dicke model.

For all other configurations, at least one emitter has a different decay rate $\sin^2(k_0 x_j)$ and Eq.~(\ref{eq: app_jump_condition}) is in general not satisfied. To illustrate this point, let us consider the specific case where all emitters are excited, $|\alpha \rangle = |E \rangle$. Then $|\beta \rangle = \hat{S} |\alpha \rangle = \sum_j \sin(k_0 x_j) |g_j \rangle$, where $|g_j \rangle$ denotes the state where only emitter $j$ is in the ground state, and Eq.~(\ref{eq: app_jump_condition}) reduces to
\begin{equation}
\label{eq: appendix_jump_condition_e}
    \sum_j \sin^3(k_0 x_j) |g_j \rangle = \beta \sum_j \sin(k_0 x_j) |g_j\rangle,
\end{equation}
for some $\beta \in \mathbb{R}$. Equation~(\ref{eq: appendix_jump_condition_e}) is only fulfilled if $\sin^2(k_0 x_j)$ is equal for all $j$, which only occurs for the generalized Dicke model. That is, applying the jump operator on an eigenstate of $\hat{H}_\mathrm{dis}$ does not necessarily result in another eigenstate of $\hat{H}_\mathrm{dis}$ for systems with $d \neq m\lambda_0/2$ and $m \in \mathbb{N}$. \\

\emph{\underline{Proposition B2}: The repeated action of the jump operator on the fully excited state necessarily populates Lindbladian dark states of $\hat{H}_\mathrm{dis} = \Gamma_B \hat{S}_B^\dagger \hat{S}_B/2$ with at least one excitation for all configurations different than the generalized Dicke model.}

\underline{Proof}: Let us first consider the situation where all emitters radiate individually,~\ie $\sin(k_0 x_j)\neq 0$ for $j \in [1,N]$. Then, acting the jump operator $N-1$ times on the excited state results in the single-excitation state
\begin{equation}
    \ket{\phi} = \hat{S}^{N-1} \ket{E} \propto \sum\limits_{j=1}^N \frac{1}{\sin(k_0 x_j)}  \ket{e_j}
\end{equation}
where $|e_j\rangle$ denotes the state where only emitter $j$ is excited. Then, $\hat{S}^\dagger \hat{S} | \phi \rangle \propto \sum_j \sin(k_0 x_j) |e_j\rangle$, and the condition $\hat{S}^\dagger \hat{S} | \phi \rangle = \phi | \phi \rangle$ is only fulfilled if $\sin^2(k_0 x_j) = \mathcal{C}$ $\forall j \in [1,N]$, where $\mathcal{C}$ is a constant. That is, $| \phi \rangle = S^{N-1} |E\rangle$ is not an eigenstate of $\hat{H}_\mathrm{dis}$ for all configurations different than the generalized Dicke model, and necessarily needs to a be superposition of at least two eigenstates of $\hat{H}_\mathrm{dis}$. Since $\hat{H}_\mathrm{dis}$ has one bright eigenstate and $N-1$ dark eigenstates (proof in Appendix~\ref{appendix: numberDarkStates}), $| \phi \rangle$ necessarily overlaps with at least one single-excitation Lindbladian state.

If $m$ emitters are dark (\ie they fulfill $\sin(k_0 x_j) =0$), they do not contribute to the decay process. Thus, applying the jump operator $N-m$ times results in the state where all bright emitters are in the ground state and the $m$ dark emitters are excited. If the jump operator is instead applied $N-m-1$ times, the system reduces to the case considered above for the $N-m$ bright emitters. Therefore, for every configuration other than the generalized Dicke system, this process directly populates at least one Lindbladian dark state containing $m+1$ excitations.

\section{Commutation of dissipative and coherent interactions}
\label{appendix: commutation}

\emph{\underline{Proposition C1}: Consider a dissipative Hamiltonian built from a single jump operator, $\hat H_\mathrm{dis} = \gamma_0 \hat{S}^\dagger \hat{S}$ with $\hat{S}=\sum_{j} s_j \hat{\sigma}_j^{ge}$ and arbitrary real coefficients $s_j$. Then $\hat H_\mathrm{dis}$ commutes with a Hermitian, excitation-number-preserving quadratic Hamiltonian $\hat H$ on the single-excitation manifold if and only if $\hat{S}$ is also an eigenoperator of $\hat{H}$. Given this, commutation on the two-excitation manifold is attained if and only if, partitioning the emitters into the bright set $B$ (those with $s_j \neq 0$) and dark set $D$ (those with $s_j = 0$), (i) $\hat H$ does not couple $B$ to $D$; and (ii) $\hat H$ restricted to $B$ is of the form $\hat H|_B = c_1\hat H_\mathrm{dis}|_B + c_2\hat N_B$ for real constants $c_1,c_2$, with $\hat N_B\equiv\sum_{j\in B}\hat\sigma_j^{eg}\hat\sigma_j^{ge}$ the bright-sector excitation number. Conditions (i)-(ii) suffice for commutation on \emph{every} higher excitation manifold as well.}

\underline{Proof:} We consider a dissipative Hamiltonian $\hat{H}_\mathrm{dis} = \gamma_0 \hat{S}^{\dagger} \hat{S} $ with a single jump operator $\hat{S} = \sum_{j=1}^N s_j \hat{\sigma}_j^{ge}$. The most general Hermitian, excitation-number-conserving quadratic coherent interaction can be written as
\begin{equation}
\label{eq: H_general}
    \hat H = \sum_{p,q=1}^N M_{pq}\,\hat\sigma_p^{eg}\hat\sigma_q^{ge},
\end{equation}
with $M$ a real symmetric $N\times N$ matrix. We keep $M$ general in what follows, and only substitute the specific form for the half-waveguide once the general result is established. Defining the operator $\hat{\sigma}^z = |e\rangle \langle e| - |g\rangle \langle g|$, we can express the commutator between the dissipative and coherent Hamiltonian as 
\begin{align}
\label{eq: general_commutator}
    \left[\hat H_\mathrm{dis},\hat H\right] = \gamma_0\sum_{j,l,m} \big[s_ls_m M_{mj} -s_js_m M_{ml}\big]\,\hat\sigma_j^{eg}\hat\sigma_m^z\hat\sigma_l^{ge}.
\end{align}

\subsubsection{Single-excitation manifold}

Restricting Eq.~(\ref{eq: general_commutator}) to the single-excitation manifold $\{|e_j\rangle\}_{j=1}^N$ gives
\begin{equation}
\label{eq: 1exc}
    \langle e_j|\left[\hat H_\mathrm{dis},\hat H\right]|e_l\rangle = \gamma_0\Big[s_j (M\vec s)_l - s_l(M\vec s)_j\Big],
\end{equation}
which vanishes for all $j,l$ if and only if $M\vec s=\lambda\vec s$ for some $\lambda\in\mathbb R$. Here, $\vec s$ corresponds to the vector with $s_j$ in index $j$, and $(M \vec s)_l = \sum_{m=1}^N M_{lm} s_m \equiv u_l$. To see this, we can consider any bright emitter $j_0$ with $s_{j_0} \neq 0$ and define $\lambda \equiv u_{j_0}/s_{j_0}$ (if no such index exists $\hat H_\mathrm{dis}=0$ trivially). Then, the vanishing condition $s_ju_l=s_lu_j$ evaluated for every $l$ gives $u_l=\lambda s_l$, or, in vector form, $M \vec s = \lambda \vec s$.

Since $M$ represents $\hat H$ on the single-excitation manifold, $M_{pq}=\langle e_p|\hat H|e_q\rangle$, and $\vec s$ represents the bright state $ | B \rangle \propto \hat S^\dagger|g\cdots g\rangle=\sum_js_j|e_j\rangle$ created by $\hat S^\dagger$ from the vacuum, the condition $M\vec s=\lambda\vec s$ is precisely the statement
\begin{equation}
\label{eq: eigenop_condition}
    \hat H\,\hat S^\dagger|g\cdots g\rangle = \lambda\,\hat S^\dagger|g\cdots g\rangle,
\end{equation}
i.e.\ that $\hat S$ is an eigenoperator of $\hat H$ within the single-excitation sector (alternatively, that the single-excitation bright state $|B\rangle$ is also an eigenstate of the coherent interactions). We impose this condition, $M\vec s=\lambda\vec s$, from here on.

\subsubsection{Two-excitation manifold}

Restricting Eq.~(\ref{eq: general_commutator}) to the two-excitation manifold, the only nonvanishing matrix elements are
\begin{align}
\label{eq: 2exc}
    \langle e_je_l|\left[\hat H_\mathrm{dis},\hat H\right]|e_je_m\rangle 
    &= \gamma_0\Big[2s_js_mM_{jl} - 2s_js_lM_{jm} \notag\\
    &\quad - s_m(M\vec s)_l + s_l (M\vec s)_m\Big],
\end{align}
with $j\neq l\neq m$. Using $M\vec s=\lambda\vec s$, the last two terms cancel exactly, reducing the condition for commutation in the two-excitation manifold as
\begin{equation}
\label{eq: 2exc_condition}
    s_j\big(s_mM_{jl} - s_lM_{jm}\big) = 0 \qquad \forall\, j\neq l\neq m.
\end{equation}

Equation~(\ref{eq: 2exc_condition}) must hold for every distinct triple $j,l,m\in[1,N]$. We differentiate the subset $B$ of emitters that are bright ($s_j \neq 0$) and the subset $D$ of emitters that are dark ($s_j = 0$).

\emph{(a) $j\in D$.} The prefactor $s_j$ vanishes, so Eq.~(\ref{eq: 2exc_condition}) imposes \emph{no constraint} on $M_{jl}$ or $M_{jm}$. 

\emph{(b) $j\in B$.} Since $s_j\neq0$, we need $s_mM_{jl}=s_lM_{jm}$ for all $l,m\neq j$.
\begin{itemize}
    \item If $m\in D$ and $l\in B$: the condition becomes $s_lM_{jm}=0$, and since $s_l\neq0$,
    \begin{equation}
    \label{eq: bright_dark_decouple}
        M_{jm}=0 \qquad \forall\, j\in B,\ m\in D.
    \end{equation}
\item If $l,m\in B$ (this requires a system with at least three bright emitters): the condition becomes $M_{jl}/s_l=M_{jm}/s_m\equiv\kappa_j$, with $\kappa_j$ independent of $l$ and $m$. Noting that we can relabel the indices in Eq.~(\ref{eq: 2exc_condition}), it follows that this condition also needs to hold when permuting indices, and in particular, $M_{lj}/s_j=M_{lm}/s_m\equiv\kappa_l$. Since $M_{jl} = M_{lj}$, it follows that $\kappa_j / s_j = \kappa_l / s_l \equiv \kappa$ and so
    \begin{equation}
    \label{eq: bright_offdiag}
        M_{jl} = \kappa\, s_js_l \qquad \forall\, j\neq l \in B.
    \end{equation}
\end{itemize}
The diagonal $M_{jj}$ for $j\in B$ follows from $M\vec s=\lambda\vec s$ together with Eqs.~(\ref{eq: bright_dark_decouple})--(\ref{eq: bright_offdiag}). Since dark sites do not contribute to $(M\vec s)_j$, we find
\begin{equation}
    \lambda s_j = M_{jj}s_j + \sum_{l\in B\setminus\{j\}}\kappa s_j s_l^2 \ \Rightarrow\ M_{jj} = \big(\lambda-\kappa|\vec s|^2\big) + \kappa s_j^2.
\end{equation}
Together with Eq.~(\ref{eq: bright_offdiag}) this gives
\begin{equation}
\label{eq: bright_block}
    M_{jl} = \kappa\, s_js_l + (\lambda-\kappa|\vec s|^2)\,\delta_{jl} \qquad \forall j,l\in B.
\end{equation}
Thus, the Hamiltonian restricted to the bright subspace commutes with the dissipative interactions if and only if it takes the form
\begin{equation}
\label{eq: bright_block_operator}
    \hat H\big|_B = \frac{\kappa}{\gamma_0}\,\hat H_\mathrm{dis}\big|_B + (\lambda-\kappa|\vec s|^2)\,\hat N_B,
\end{equation}
for $\lambda, \kappa \in \mathbb R$ and where $\hat N_B \equiv \sum_{j \in B} \hat{\sigma}_j^{eg} \hat{\sigma}_j^{ge}$ is the excitation-number operator restricted to the bright subset $B$. The coherent Hamiltonian restricted to the bright emitters, once both the one- and two-excitation commutation conditions are imposed, must be a real multiple of the dissipative Hamiltonian's bright block, plus a term proportional to the bright-sector excitation-number operator (the latter being trivial in the sense that it acts as a constant or identity operator within each fixed-excitation-number sector). It is worth noting that the derivation assumed at least three bright emitters: if the system has only one or two bright emitters, then commutation of the coherent and dissipative Hamiltonian within the bright subspace is ensured by the single-excitation condition.

Finally, taking case (a), it follows that the Hamiltonian restricted to the dark subspace ($\hat{H}|_{D}$) is completely unconstrained and can take any form that preserves Hermiticity. From the first sub-case of (b), however, it follows that commutation between the coherent and dissipative Hamiltonians requires no coherent interaction between bright and dark emitters. In other words, $\hat H$ must be block-diagonal in the bright/dark partition ($\hat{H}|_{BD}=0$).

\subsubsection{Extension to all excitation manifolds}

Since commutation for the two-excitation manifold requires $\hat H|_{BD}=0$, the Hilbert space factorizes as $\mathcal H=\mathcal H_B\otimes\mathcal H_D$ and $\hat H=\hat H\big|_B\otimes\hat{I}\big|_D+\hat{I}\big|_B\otimes\hat H\big|_D$. As $\hat S$ involves only bright sites, $\hat H_\mathrm{dis}=\hat H_{\mathrm{dis}}\big|_B\otimes\hat{I}\big|_D$ acts trivially on the dark sector, so
\begin{equation}
    [\hat H_\mathrm{dis},\hat H] = \left[\hat H_{\mathrm{dis}}\big|_B,\hat H\big|_B \right]\otimes\hat{I}\big|_D,
\end{equation}
independent of $\hat H\big|_D$ and of which excitation manifold of the full system is considered. The problem thus reduces to whether $\hat H\big|_B=c_1\hat H_{\mathrm{dis}}\big|_B+c_2\hat N_B$ (with $c_1,c_2 \in \mathbb{R}$) commutes with $\hat H_{\mathrm{dis}}\big|_B$ on the bright subsystem alone. Indeed, it trivially does on every excitation manifold, since $\hat H_{\mathrm{dis}}\big|_B$ commutes with itself and $\hat N_B$ commutes with any excitation-number-conserving operator. No further restriction on $c_1$, $c_2$ or $\hat H\big|_D$ arises beyond what the one- and two-excitation conditions already fixed, completing the proof of Proposition C1. \\

\emph{\underline{Proposition C2}: For the coherent and dissipative couplings of the half-waveguide [Eq.~(\ref{eq: coh_dis_int})], the conditions in Proposition C1 are satisfied only when the dark emitters occupy the left-most positions in the chain, closest to the mirror, and every bright emitter except the left-most one sits at an equivalent point of the standing wave modulo $\lambda_0/2$ (the generalized Dicke model), provided that the left-most bright emitter is further detuned by $\Delta_\mathrm{opt}$ [Eq.~(\ref{eq: optimal_detuning_2atoms})] from the rest of the bright emitters.}

\underline{Proof:} It is instructive to express the dipole-dipole couplings in Eq.~(\ref{eq: coh_dis_int}) as
\begin{align}
\label{eq: J_closed}
    J_{jl} &= -\gamma_0\sin(k_0x_<)\cos(k_0x_>),\\
\label{eq: Gamma_closed}
    \Gamma_{jl} &= 2\gamma_0\sin(k_0x_j)\sin(k_0x_l),
\end{align}
where we have defined $x_>\equiv\max(x_j,x_l)$ and $x_<\equiv\min(x_j,x_l)$. Using Eq.~(\ref{eq: Gamma_closed}), the dissipative Hamiltonian $\hat{H}_\mathrm{dis} = \sum_{j,l} \Gamma_{jl} \hat{\sigma}_j^{eg} \hat{\sigma}_l^{ge} / 2$ can be written in the form of Proposition C1, $\hat{H}_\mathrm{dis} = \gamma_0 \hat{S}^\dagger S$, with $\hat S = \sum_j \sin (k_0 x_j)\,\hat\sigma_j^{ge}$, such that $s_j \equiv \sin (k_0 x_j)$. \\

\emph{Dark emitters decouple from bright ones only if positioned closer to the mirror.} Equation~(\ref{eq: J_closed}) depends on $\sin(k_0x_<)$, so a dark emitter ($s_j=0$) gives $J_{jl}=0$ only if $x_j<x_l$. Otherwise, $J_{jl}=-\gamma_0 s_l\cos(k_0x_j)\neq0$ despite $s_j=0$. Full decoupling from every bright emitter, as required by Eq.~(\ref{eq: bright_dark_decouple}), therefore demands the dark emitter lie to the left of all bright ones (i.e., closer to the mirror). Thus, a set of dark emitters is only decoupled from the bright emitters if it consists of the left-most emitters in the chain. Note also that Eq.~(\ref{eq: J_closed}) implies dark emitters are automatically decoupled from one another, independently of their positions.

\emph{The bright block realizes Eq.~(\ref{eq: bright_block_operator}) only for special geometries.} For a bright pair, Eq.~(\ref{eq: J_closed}) gives $J_{jl}=-\gamma_0\sin(k_0x_<)\cos(k_0x_>)$, which must equal the corresponding off-diagonal term in Eq.~(\ref{eq: bright_offdiag}), $\kappa\,s_js_l=\kappa\sin(k_0x_<)\sin(k_0x_>)$. This requires
\begin{equation}
\label{eq: dicke_realizability}
    \cot(k_0x_>) = -\kappa/\gamma_0 \quad \text{for every bright pair,}
\end{equation}
with the \emph{same} $\kappa$ throughout. Ordering the bright emitters by position, $x_{(1)}<x_{(2)}<\dots<x_{(N_B)}$, only $x_{(2)},\dots,x_{(N_B)}$ ever appear as $x_>$ in Eq.~(\ref{eq: dicke_realizability}). The left-most bright emitter $x_{(1)}$ never does, and is therefore left unconstrained. Equation~(\ref{eq: dicke_realizability}) can hold simultaneously for all bright pairs only if $x_{(2)},\dots,x_{(N_B)}$ coincide modulo $\lambda_0/2$, that is, if all but the first emitter sit at the same point of the standing wave. In this case, the bright block reduces to the generalized Dicke model. Away from this fine-tuned geometry, the off-diagonal part of $\hat H|_B$ cannot be brought into the form~(\ref{eq: bright_block_operator}) by any choice of detunings $\Delta_j$, since $\Delta_j$ only adjust the diagonal terms of $\hat H|_B$.

Attaining commutation within the bright block also requires Eq.~(\ref{eq: bright_block}) to be fulfilled on the diagonal. This implies $M_{jj} = \kappa \sin^2(k_0x_j) \equiv -\gamma_0\sin^2(k_0x_j)\cos(k_0x_{(2)})/\sin(k_0x_{(2)})$, up to an arbitrary global energy shift captured by the term proportional to $\hat N_B$. For every bright emitter at positions $x_{(2)},\dots,x_{(N_B)}$, this is automatically satisfied by the physical Lamb shift alone, $M_{jj}=J_{jj}$. For the left-most bright emitter $l \equiv (1)$, however, $M_{ll}\neq J_{ll}$ in general, and a detuning is required to fulfill Eq.~(\ref{eq: bright_block}) on the diagonal. Writing $M_{ll}=J_{ll}+\Delta_{l}$, the required detuning is exactly $\Delta_\mathrm{opt}$ of Eq.~(\ref{eq: optimal_detuning_2atoms}), with the second emitter's position there replaced by any one of $x_{(2)},\dots,x_{(N_B)}$. Commutation within the bright block therefore further requires detuning the left-most bright emitter from the rest by $\Delta_\mathrm{opt}$, which vanishes if its position also happens to coincide modulo $\lambda_0/2$. Note that an additional, uniform detuning can always be applied to every bright emitter simultaneously, since this corresponds to a term proportional to $\hat N_B$ in $\hat H$.

\emph{Conclusion.} Full commutation between $\hat H$ and $\hat H_\mathrm{dis}$ therefore requires (i) all dark emitters placed at the left-most positions of the chain, closest to the mirror, and (ii) all bright emitters except the left-most one to be positioned at equivalent points of the standing wave (generalized Dicke model), with the optimal detuning $\Delta_\mathrm{opt}$ applied between the first bright emitter and the rest.

\section{Stabilizing two-excitation dark states in systems of four bright emitters}
\label{Appendix: 4atoms_2exc}
Let us consider a general two-excitation Lindbladian dark state of a system of four bright emitters, which can be written as $|\psi_\mathcal{D}^{(2)} \rangle = \beta  |D_{12,34} \rangle + \alpha |D_{13,24} \rangle$ with $|D_{12,34} \rangle$ and $|D_{13,24} \rangle$ defined in Eq.~\eqref{eq: darkstate_2exc_separable}. In the basis $\{ |eegg\rangle,|egeg\rangle,|egge\rangle,|geeg\rangle,|gege\rangle,|ggee\rangle \} $, we can write this state as the column vector
\begin{widetext}
\begin{equation}
    |\psi_\mathcal{D}^{(2)} \rangle=
    \begin{pmatrix}
    \psi_1 \\ \psi_2 \\ \psi_3 \\ \psi_4 \\ \psi_5 \\ \psi_6
    \end{pmatrix} = \beta
    \begin{pmatrix}
    0 \\ \sin(k_0 x_2)  \sin(k_0 x_4)  \\ -\sin(k_0 x_2)  \sin(k_0 x_3)  \\  - \sin(k_0 x_1)  \sin(k_0 x_4)   \\ \sin(k_0 x_1)  \sin(k_0 x_3)   \\ 0
    \end{pmatrix} + \alpha
    \begin{pmatrix}
    \sin(k_0 x_3)  \sin(k_0 x_4)  \\ 0 \\ -\sin(k_0 x_2)  \sin(k_0 x_3) \\  - \sin(k_0 x_1)  \sin(k_0 x_4)   \\ 0  \\ \sin(k_0 x_1)  \sin(k_0 x_2) 
    \end{pmatrix}.
\end{equation}

Then, equation $\hat{H} |\psi_\mathcal{D}^{(2)} \rangle = E_\mathcal{D} |\psi_\mathcal{D}^{(2)} \rangle$ results in a linear system of six equations and the four free parameters or unknowns $\vec{u} = [\Delta_2,\Delta_3,\Delta_4,E_\mathcal{D}]^T$, 
\begin{equation}
\label{eq: system_equations_4emitters_2exc}
    \mathbf{M} \vec{u}
    = -  \mathbf{J}_2 |\psi_D^{(2)} \rangle,
\end{equation}
where $\Delta_j$ is the detuning of emitter $j$ from the resonance frequency $\omega_0$.
Here, the $6 \times 4$ matrix $\mathbf{M}$ takes the form 
\begin{equation}
    \mathbf{M} = \begin{pmatrix}
    \psi_1 & 0 & 0 & -\psi_1 \\
    0 & \psi_2 & 0 & -\psi_2 \\
    0 & 0 & \psi_3 & -\psi_3 \\
    \psi_4 & \psi_4 & 0 & -\psi_4 \\
    \psi_5 & 0 & \psi_5 & -\psi_5 \\
    0 & \psi_6 & \psi_6 & -\psi_6 \\
    \end{pmatrix},
\end{equation}
and $\mathbf{J}_2$ corresponds to coherent interactions induced by the half-waveguide between the basis states of the two-excitation manifold,
\begin{equation}
    \mathbf{J}_2 \!= \!\begin{pmatrix}
    J_{11}\!+\!J_{22} & J_{23} & J_{24} & J_{13} & J_{14} & 0 \\
    J_{23} & J_{11}\!+\!J_{33} & J_{34} & J_{12} & 0 & J_{14} \\
    J_{24} & J_{34} & J_{11}\!+\!J_{44} & 0 & J_{12} & J_{13} \\
    J_{13} & J_{12} & 0 & J_{22}\!+\!J_{33} &  J_{34} & J_{24} \\
    J_{14} & 0 & J_{12} & J_{34} & J_{22}\!+\!J_{44} & J_{23} \\
    0 & J_{14} & J_{13} & J_{24} & J_{23} & J_{33}\!+\!J_{44} 
    \end{pmatrix}.
\end{equation}
\end{widetext}
The general solution to Eq.~(\ref{eq: system_equations_4emitters_2exc}) is given by the Moore--Penrose pseudoinverse, $\vec{u} = -\mathbf{M}^+ \mathbf{J}_2 |\psi_D^{(2)} \rangle$, which minimizes $\|\mathbf{M} \vec{u}+ \mathbf{J}_2 |\psi_D^{(2)} \rangle\|$. This reduces to the ordinary least-squares expression $\vec u = -(\mathbf M^T\mathbf M)^{-1}\mathbf M^T\mathbf J_2|\psi_D^{(2)}\rangle$ whenever $\mathbf M$ has full column rank, but remains well defined even when $\mathbf M^T\mathbf M$ is singular (as happens for $\alpha=0$, $\beta=0$ or $\alpha=-\beta$), in which case it returns the smallest-norm solution among all those minimizing the distance. Interestingly, this minimization has a clear physical interpretation. Noting that non-perfect Lindbladian dark states fulfill $\hat{H} |\psi_{D}^{(2)}\rangle = E_D |\psi_{D}^{(2)}\rangle + J | \!\perp\! \psi_{D}^{(2)}\rangle$, with $|\!\perp\! \psi_{D}^{(2)}\rangle$ a normalized state orthogonal to $|\psi_{D}^{(2)}\rangle$, it follows that the minimized distance is precisely $|J|$. The pseudoinverse solution therefore provides the detuning pattern that minimizes leakage (that is, the coupling to states other than $|\psi_D^{(2)}\rangle$ caused by $\hat{H}$).

Direct interactions between qubits independently of the waveguide increase the number of free parameters. Introducing an external coupling between emitters $1$ and $2$, of magnitude $\tilde{J}_{12}$, and between emitters $2$ and $3$, of magnitude $\tilde{J}_{23}$, the new vector of six unknowns is $\vec{u}_2 = [\Delta_2,\Delta_3,\Delta_4,E_\mathcal{D},\tilde{J}_{12},\tilde{J}_{23}]^T$ and the linear system of equations reads
\begin{equation}
\label{eq: liner_syst_couplers}
    \mathbf{M}_2 \vec{u}_2
    = -  \mathbf{J}_2 |\psi_\mathcal{D}^{(2)} \rangle,
\end{equation}
where $\mathbf{M}_2$ is now the $6 \times 6$ matrix 
\begin{equation}
    \mathbf{M}_2 = \begin{pmatrix}
    \psi_1 & 0 & 0 & -\psi_1 & 0 & \psi_2 \\
    0 & \psi_2 & 0 & -\psi_2 & \psi_4 & \psi_1  \\
    0 & 0 & \psi_3 & -\psi_3 & \psi_5 & 0 \\
    \psi_4 & \psi_4 & 0 & -\psi_4 & \psi_2 & 0 \\
    \psi_5 & 0 & \psi_5 & -\psi_5 & \psi_3 & \psi_6 \\
    0 & \psi_6 & \psi_6 & -\psi_6 & 0 & \psi_5 \\
    \end{pmatrix},
\end{equation}
The solution of Eq.~(\ref{eq: liner_syst_couplers}) is given by $\vec{u}_2 = -\mathbf{M}_2^{+} \mathbf{J}_2 |\psi_D^{(2)} \rangle$. For $\alpha = 0$, $\beta = 0$, or $\alpha = -\beta$, $\mathbf{M}_2$ is singular, and $\vec{u}_2$ corresponds to the least-squares solution of Eq.~(\ref{eq: liner_syst_couplers}). For all other combinations of $\alpha$ and $\beta$, $\mathbf{M}_2$ is invertible, and $\vec{u}_2 = -\mathbf{M}_2^{-1} \mathbf{J}_2 |\psi_D^{(2)} \rangle$ gives the detunings and external couplings required to make $|\psi_\mathcal{D}^{(2)} \rangle$ a perfect dark state of the full master equation. Importantly, not all choices of couplings $\tilde{J}_{jl}$ yield a matrix $\mathbf{M}_2$ that is generically invertible and, hence, allow for a perfect dark state. For example, this is not the case when one coupler connects the first and second emitters ($\tilde{J}_{12}$) and another connects the third and fourth emitters ($\tilde{J}_{34}$).

\section{Preparation of localized dark states through heralded dissipation}
\label{Appendix: heralded_dissipation}

We consider a driving field that incoherently excites a subset $\mathcal{E}$ of emitters, $| \psi_\mathrm{incoh} \rangle = \prod_{j \in \mathcal{E}} \hat{\sigma}_j^{eg} |G\rangle$. This can be achieved either by externally driving all emitters in $\mathcal{E}$, or by applying a drive through the waveguide and simultaneously detuning all emitters not in $\mathcal{E}$. Note that the driving field needs to be strong such that the light-induced interactions are negligible during the excitation process. While $| \psi_\mathrm{incoh} \rangle$ is not a dark state (\ie $\hat{S}_B | \psi_\mathrm{incoh} \rangle \neq 0$), it can have a large overlap with dark states of the system or decay with high probability into them. Then, measuring the photon count at the output of the waveguide projects the state into the desired dark states with finite probability and allows for their preparation via heralded dissipation.

To illustrate this process, we consider again a two-emitter system with the single-excitation Lindbladian dark state $|D\rangle = [ \sin(k_0x_2) |e_1\rangle - \sin(k_0x_1) |e_2\rangle] / \sqrt{\sin(k_0x_1)^2 +\sin(k_0x_2)^2}$. Its overlap with the incoherent state $| \psi_\mathrm{incoh} \rangle = |e_1 \rangle$ is finite and equal to $p_D = |\langle  \psi_\mathrm{incoh} | D \rangle |^2 = \sin^2(k_0 x_2) /(\sin^2(k_0 x_1)  +\sin^2(k_0 x_2) )$. If the optimal detuning in Eq.~(\ref{eq: optimal_detuning_2atoms}) is applied, $|D\rangle$ is additionally a dark state of the full master equation. With probability $1-p_D$, the excitation decays after a short time, a photon is measured at the output of the waveguide and the system of emitters collapses to the ground state $|G\rangle$. If no photon is measured, however, the probability to be in $|D\rangle$ for that specific trajectory or realization of the experiment rapidly increases with time, and eventually results in the preparation of state $|D\rangle$. This occurs with probability $p_D$, which is maximized for dissimilar decay rates of the individual emitters.

For multiple emitters, the single-excitation bright state can be expressed as $|B\rangle = \sum_j \sin(k_0 x_j)  \hat{\sigma}_j^{eg} |G\rangle / \sqrt{\sum_j \sin^2(k_0 x_j)}$. Defining the operator $\hat{S}_2^\dagger = \sum_{j \neq d} \sin(k_0 x_j)  \hat{\sigma}_j^{eg} / \sqrt{\sum_{j \neq d} \sin^2(k_0 x_j)} $, we can express the bright state as 
\begin{equation}
    |B\rangle = \frac{ \sin(k_0 x_d)  \hat{\sigma}_d^{eg} + \sqrt{\sum_{j\neq d} \sin^2(k_0 x_j) } \hat{S}_2^\dagger }{\sqrt{\sum_j \sin^2(k_0 x_j) }} |G\rangle. 
\end{equation}
It is then straightforward to construct the single-excitation localized dark state orthogonal to $|B \rangle$,
\begin{equation}
\label{eq: localized_darkstate}
    |D_\mathrm{loc}\rangle = \frac{ \sqrt{\sum_{j\neq d} \sin^2(k_0 x_j) } \hat{\sigma}_d^{eg} - \sin(k_0 x_d)   \hat{S}_2^\dagger }{\sqrt{\sum_j \sin^2(k_0 x_j) }} |G\rangle,
\end{equation}
which contains a significant fraction of the excitation, namely $p_D^{(\mathrm{loc})} = 1-\sin^2(k_0 x_d)/(\sum_j \sin^2(k_0 x_j))$, at emitter $d$. In other words, incoherently exciting emitter $d$ results in a state $| \psi_\mathrm{incoh} \rangle = |e_d \rangle$ that has large overlap $p_D^{(\mathrm{loc})}$ with the localized dark state $|D_\mathrm{loc}\rangle$. For interactions with permutational symmetry ($d=m \lambda_0/2$ with $m \in \mathbb{N}$), the spontaneous decay rate of all emitters is identical (\ie $\sin(k_0x_j)$ is independent of $j$), $|D_\mathrm{loc} \rangle = (\sqrt{N-1} \hat{\sigma}_d^{eg} -  \hat{S}_2^\dagger) |G\rangle / \sqrt{N}$ reduces to that introduced in Ref.~\cite{Holzinger22} and we obtain $p_D^{(\mathrm{loc})} = 1-1/N$. For other configurations, each emitter $j$ can in general have a different decay rate $\Gamma_{jj} \propto \sin^2(k_0 x_j)$. Then, exciting the emitter $d$ with the smallest decay rate results in a larger overlap with the desired dark state $|D_\mathrm{loc} \rangle$. This increases the efficiency of the preparation protocol, even if the number of emitters remains moderate. Note also that these results can be readily generalized to localized multiexcitation dark states.

\section{Generating dark states with weak classical driving fields}
\label{Appendix: Generating_dark_classical_drive}

\subsection{Adiabatic elimination for system with two emitters}

\begin{figure}[t!]
    \includegraphics[width=\columnwidth]{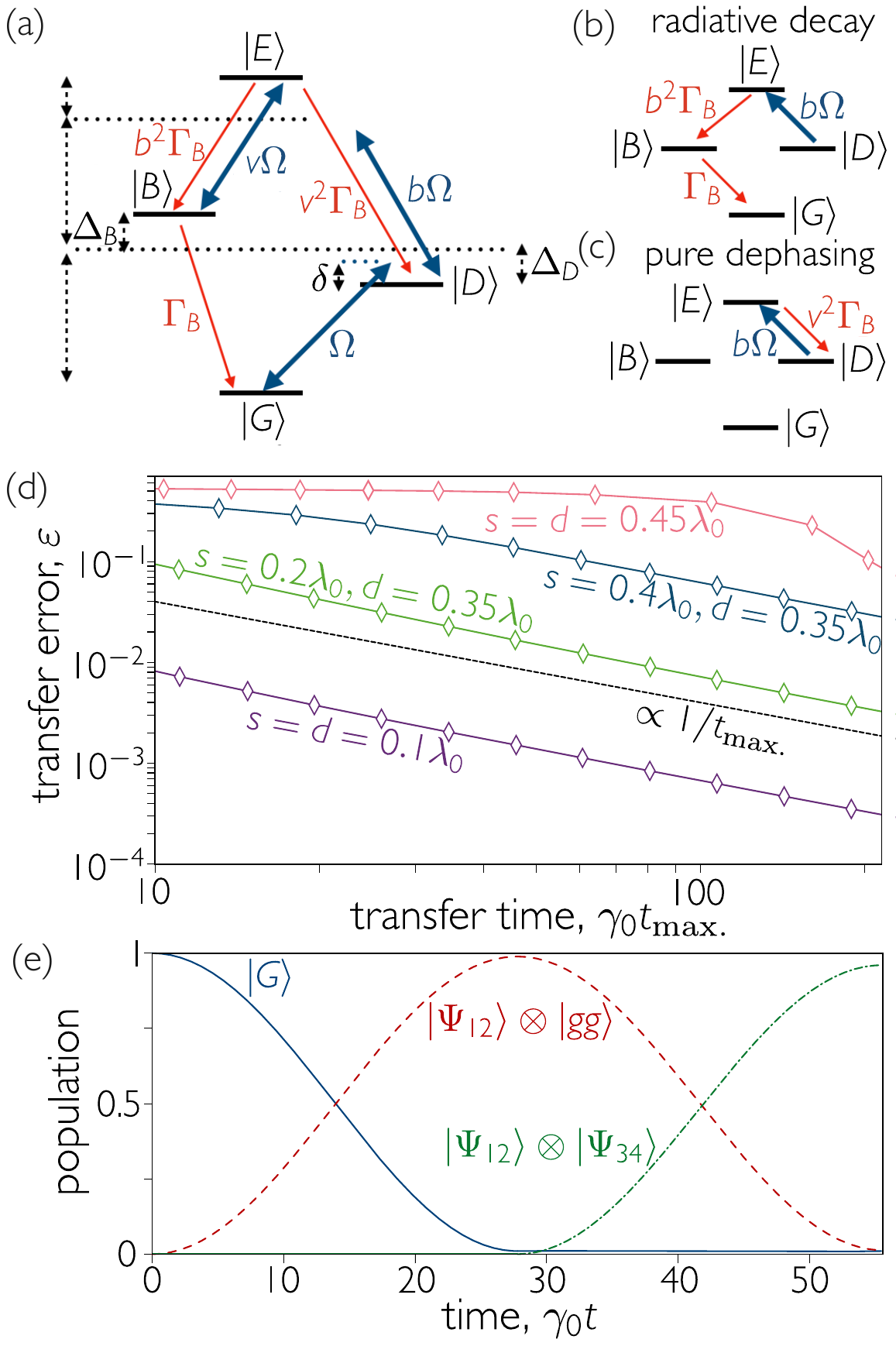}
    \caption{(a) Collective states of a two-emitter system at the optimal detuning. $|B\rangle$ and $|D\rangle$ denote the single-excitation bright and dark states, respectively. The possible decay paths are depicted by red arrows, while the action of the drive with strength $\Omega$ is illustrated by the blue double arrows. (b)-(c) Mechanism leading to (b) radiative decay and (c) pure dephasing (as well as their corresponding energy shifts) for the $|D\rangle \leftrightarrow |G \rangle$ transition after adiabatically eliminating the radiating states $|B\rangle$ and $|E\rangle$. (d) Scaling of the transfer error $\epsilon = 1 - p_D$ for a system with four emitters in various configurations as a function of transfer time $t_\mathrm{max}$. The black dashed line corresponds to a scaling of $\epsilon \sim t_\mathrm{max}^{-1}$ and serves as a guide to the eye. (e) Preparation scheme of the separable four-emitter two-excitation dark state $| D_{12,34} \rangle = | D_{12} \rangle \otimes | D_{34} \rangle$ for $s=\lambda_0/4$ and $d=0.15\lambda_0$ and subsystem detuning $\Delta_{\mathrm{off}} = 10\gamma_0$. Populations in the (blue solid line) total ground state $|G\rangle = | gggg\rangle $, (red dashed) the single-excitation dark state $|D_{12} \rangle \otimes |gg\rangle$, and (green dot-dashed) the two-excitation dark state $| D_{12,34} \rangle$.}
    \label{fig: adiabatic_2emitters}
\end{figure}

We consider a system of two-emitters as depicted in Fig.~\ref{fig: adiabatic_2emitters}. Applying the optimal detuning $\Delta_1=-\Delta_2=\Delta_\mathrm{opt}/2$ given in Eq.~(\ref{eq: optimal_detuning_2atoms}), the eigenstates of the coherent and dissipative interactions are the ground state $|G \rangle$, the fully excited state $|E \rangle$ and the single-excitation bright $|B\rangle$ and dark $|D\rangle$ states given by Eqs.~(\ref{eq: 2atoms_brightstate}) and~(\ref{eq: 2atoms_darkstate}), respectively. Writing the jump operator as $\hat{S}_B = |G\rangle \langle B| + b |B\rangle \langle E| + v |D\rangle \langle E|$, where we have defined 
\begin{subequations}
    \begin{align}
    b \equiv \frac{2\sin(k_0 x_1) \sin(k_0 x_2) }{\sin^2(k_0 x_1) +\sin^2(k_0 x_2) },\\
    v \equiv \frac{\sin^2(k_0 x_2) - \sin^2(k_0 x_1)}{\sin^2(k_0 x_1) +\sin^2(k_0 x_2)},
    \end{align}
\end{subequations}
we can readily compute the dissipative transition rates between the states, $\Gamma_{i \rightarrow f} = \Gamma_B |\langle f|\hat{S} |i\rangle|^2$. The only non-zero transition rates are $\Gamma_{E\to B} = b^2 \Gamma_B$, $\Gamma_{E\to D} = v^2 \Gamma_B$ and $\Gamma_{B\to G} = \Gamma_B$. Additionally, the energy shifts of the four states, resulting from the combined effects of the waveguide-mediated coherent interactions and the optimal detuning pattern, are given by
\begin{subequations}
    \begin{align}
    \Delta_G&=0 \\ \Delta_E &= J_{11}+J_{22} \\ \Delta_B &= (J_{11} +J_{22})/2 - v(J_{11} + \Delta_\mathrm{opt} -J_{22} )/2 + bJ_{12} \\ \Delta_D &= (J_{11} +J_{22})/2 + v(J_{11} + \Delta_\mathrm{opt} -J_{22} )/2 - bJ_{12}.
    \end{align}
\end{subequations}

We apply the driving Hamiltonian in Eq.~(\ref{eq: drive_dark_1exc_2at}), which selectively couples the ground state $|G\rangle$ to the dark state $|D\rangle$ with Rabi frequency $\Omega$. In the frame rotating at the drive frequency $\omega_L$, the Hamiltonian can be expressed in the eigenbasis of the undriven system as
\begin{widetext}
\begin{align}
    \hat{H} &= -\delta |D\rangle \langle D| - \tilde{\Delta}_B |B\rangle \langle B| - \tilde{\Delta}_E |E\rangle \langle E| + \Omega ( |G\rangle \langle D| + |D\rangle \langle G| )
    - b \Omega ( |D\rangle \langle E| + |E\rangle \langle D| )  + v \Omega  ( |B\rangle \langle E| + |E\rangle \langle B| ),
\end{align}
\end{widetext}
where $\tilde{\Delta}_B = \omega_L - \omega_0 - \Delta_B$, $\tilde{\Delta}_E =2 \omega_L - 2 \omega_0 -\Delta_E$ and $\delta = \omega_L - \omega_0 - \Delta_D$ is the detuning of the driving field from the resonance frequency of the dark state.

From the master equation~(\ref{eq: Master_equation}), we compute the equations of motion for the population in state $|A\rangle$, $p_A = \langle A | \hat{\rho} | A\rangle$, and for the coherences between states $|A\rangle$ and $|B\rangle$, $c_{AB}= \langle A | \hat{\rho} | B\rangle$, 
\begin{widetext}
\begin{subequations}
\begin{align}
\label{eq: app_adiab_e}
    \dot{p}_E &= -\Gamma_B p_E - i v \Omega  (c_{BE}-c_{EB}) + i b \Omega (c_{DE}-c_{ED}),\\
\label{eq: app_adiab_b}
    \dot{p}_B &= -\Gamma_B p_B + b^2 \Gamma_B p_E + i v\Omega  (c_{BE}-c_{EB}),  \\
\label{eq: app_adiab_d}
    \dot{p}_D &= v^2 \Gamma_B p_E - i b\Omega  (c_{DE}-c_{ED}) -i \Omega  (c_{GD}-c_{DG}), \\
\label{eq: app_adiab_g}
    \dot{p}_G &= \Gamma_B p_B +i \Omega  (c_{GD}-c_{DG}), \\
\label{eq: app_adiab_eb}
    \dot{c}_{EB} &= (-\Gamma_B + i (\tilde{\Delta}_E - \tilde{\Delta}_B)) c_{EB} - i v\Omega (p_B -p_E) + i b\Omega  c_{DB},  \\
\label{eq: app_adiab_ed}
    \dot{c}_{ED} &= (-\frac{\Gamma_B}{2} + i (\tilde{\Delta}_E - \delta)) c_{ED} + i b \Omega (p_D -p_E) - i v\Omega c_{BD} + i \Omega c_{EG},  \\
\label{eq: app_adiab_eg}
    \dot{c}_{EG} &= (-\frac{\Gamma_B}{2} + i \tilde{\Delta}_E ) c_{EG} + i \Omega c_{ED} - i v\Omega  c_{BG} + i b\Omega  c_{DG},  \\
\label{eq: app_adiab_bd}
    \dot{c}_{BD} &= (-\frac{\Gamma_B}{2} + i (\tilde{\Delta}_B - \delta)) c_{BD} + bv\Gamma_B p_E + i \Omega c_{BG} - i b\Omega  c_{BE} - i v\Omega  c_{ED},  \\
    \dot{c}_{BG} &= (-\frac{\Gamma_B}{2} + i \tilde{\Delta}_B ) c_{BG} + b \Gamma_B c_{EB} + i \Omega c_{BD} - i v\Omega  c_{EG}, \\
\label{eq: app_adiab_dg}
    \dot{c}_{DG} &=  i \delta c_{DG} + v \Gamma_B c_{EB} + i \Omega (p_D - p_G) + i b\Omega  c_{EG}. 
\end{align}
\end{subequations}
\end{widetext}

If the drive is weak and near on-resonance with the $|G\rangle \leftrightarrow |D\rangle$ transition, such that $\Omega,\delta \ll \Gamma_B,\tilde{\Delta}_E$, $|E\rangle$ and $|B\rangle$ evolve quickly and can be adiabatically eliminated. For that, we set the derivative of all populations and coherences describing states $|E\rangle$ and $|B\rangle$ to zero. Keeping only terms up to order $\Omega^2/\Gamma_B^2$, we obtain  
\begin{subequations}
    \begin{align}
        \Omega c_{ED} &\approx b \delta_p p_D + i b \gamma_p p_D/2 \\\Omega c_{EG} &\approx b \delta_p c_{DG} + i b \gamma_p c_{DG}/2 \\ \Gamma_B p_E &\approx b^2\gamma_p  p_D \\ \Gamma_B p_B &\approx b^4\gamma_p  p_D \\ c_{EB} &= 0,
    \end{align}
\end{subequations}
where we have defined the damping rate $\gamma_p$ and the shift $\delta_p$ as 
\begin{align}
\label{eq: adiab_decay}
    \gamma_p &= \frac{\Omega^2}{(\Delta_E \! -\!2\Delta_D)^2 + (\Gamma_B/2)^2} \Gamma_B, \\
\label{eq: adiab_shift}
    \delta_p &= \frac{\Omega^2}{(\Delta_E\!-\!2\Delta_D)^2 + (\Gamma_B/2)^2} (\Delta_E\!-\!2\Delta_D). 
\end{align}
Plugging these expressions in Eqs.~(\ref{eq: app_adiab_d}), (\ref{eq: app_adiab_g}) and (\ref{eq: app_adiab_dg}), we finally obtain the dynamics for the subsystem composed of $|G\rangle$ and $|D\rangle$ as
\begin{subequations}
\label{eq: effective_dynamics_2atoms_dark}
\begin{align}
\label{eq: adiabatic_populations}
    \dot{p}_D &= - \dot{p}_G = -\gamma_r p_D - i \Omega (c_{GD} - c_{DG}), \\
\label{eq: adiabatic_coherence}
    \dot{c}_{DG} &= \left( i ( \delta + \delta_r +  \delta_d ) - \frac{\gamma_r+\gamma_d}{2} \right)  c_{DG} + i \Omega (p_D-p_G).
\end{align}
\end{subequations}
Here, $p_D$ and $p_G$ respectively denote the population in $|D\rangle$ and $|G\rangle$, and $c_{DG}$ corresponds to the coherence between both states. Equation~(\ref{eq: effective_dynamics_2atoms_dark}) describes Rabi oscillations between the ground state $|G\rangle$ and the single-excitation dark state $|D\rangle$. Additionally, the undesired coupling between $\ket{D}$ and $\ket{E}$ results in two additional processes. First, an effective decay from $|D\rangle$ to $|G\rangle$, which occurs due to population of the excited state, followed by decay to the single-excitation bright state $|B\rangle$, and finally to $|G\rangle$. This process, illustrated in Fig.~\ref{fig: adiabatic_2emitters}(b), occurs at a rate $\gamma_r = b^4 \gamma_p$. Second, an effective dephasing of the single-excitation dark state $|D\rangle$, which instead occurs due to decay from $\ket{E}$ to $\ket{D}$. This process is illustrated in Fig.~\ref{fig: adiabatic_2emitters}(c) and occurs at a rate $\gamma_d = b^2 v^2 \gamma_p$. Additionally, each process shifts the resonance frequency of $\ket{D}$ by $\delta_r = b^4 \delta_p$ and $\delta_d = b^2 v^2 \delta_p$, respectively. Notably, the emerging radiative decay and dephasing processes are weak compared to the strength of the drive (\ie $\gamma_p,\delta_p \ll \Omega$) if either $\Gamma_B \gg \Omega$  or $|\Delta_E-2\Delta_D| \gg \Omega$. The former case corresponds to suppression of the process $|D\rangle$ to $|E\rangle$ by the Zeno effect~\cite{Facchi02}. The latter case corresponds to strong off-resonance of the transition from $|D\rangle$ to $|E\rangle$ for drive on resonance with the $|G\rangle \leftrightarrow |D\rangle$ transition, which leads to a photon-blockade effect~\cite{Cidrim20,Williamson20PRL}. 

Solving Eq.~(\ref{eq: effective_dynamics_2atoms_dark}) for $\delta = -\delta_r-\delta_d = -b^2 \delta_p$ (\ie for resonant drive) and for a system initially in the ground state, we obtain the population in the dark state $|D\rangle$ as a function of time
\begin{widetext}
\begin{align}
    p_D&=\frac{1}{2(1+ \frac{\gamma_r (\gamma_r+\gamma_d)}{8 \Omega^2} )} \left( 1 - \cos (2 \Omega_\mathrm{eff} t) e^{-\frac{3 \gamma_r + \gamma_d}{4} t}\right) \nonumber - \frac{1}{2(1+ \frac{\gamma_r (\gamma_r+\gamma_d)}{8 \Omega^2} )} \frac{3 \gamma_r + \gamma_d}{8 \Omega_\mathrm{eff}} \sin (2 \Omega_\mathrm{eff} t) e^{-\frac{3 \gamma_r + \gamma_d}{4} t},
\end{align}
\end{widetext}
where we have introduced the generalized Rabi frequency $\Omega_\mathrm{eff} = \sqrt{\Omega^2 -(\gamma_r -\gamma_d)^2/64}$. The maximum population in the dark state is attained at $t_\mathrm{max} \approx \pi/2\Omega_\mathrm{eff}$ and is given by
\begin{equation}
\label{eq: adiab_maximum_pop_SI}
    p_D^{(\mathrm{max})} \approx \frac{1}{2 \left( 1+ \frac{\gamma_r (\gamma_r+\gamma_d)}{8 \Omega^2} \right)} \left( 1 + e^{-\frac{3 \gamma_r + \gamma_d}{4} \frac{\pi}{2\Omega_\mathrm{eff}} }\right), 
\end{equation}
from which one can readily derive the error $\epsilon = 1 - p_D^{(\mathrm{max})}$ of the dark state preparation protocol in Eq.~(\ref{eq: scaling_error}).

\subsection{Additional numerical data}

Single-excitation dark states of systems with more than two emitters can also be prepared by applying a weak drive that matches the structure of the dark state, as discussed in Section~\ref{section: dark_state_generation_classical}. As occurred for the case of two emitters, the error of the preparation protocol scales as $\epsilon \propto t_\mathrm{max}^{-1}$, as numerically shown in Fig.~\ref{fig: adiabatic_2emitters}(d). The error is smaller for configurations where two-excitation states coupled by the drive are bright and far off-resonant.

In Fig.~\ref{fig: adiabatic_2emitters}(e), we show the preparation protocol for a two-excitation separable dark state of a four-emitter system, $| D_{12,34} \rangle =| D_{12} \rangle \otimes | D_{34} \rangle $ given in Eq.~(\ref{eq: darkstate_2exc_separable}), as discussed in Section~\ref{section: dark_state_generation_classical}. We obtain high efficiencies by sequentially driving each subsystem.

\section{Robustness of dark-state generation via few-photon absorption}
\label{Appendix: few_photon_absorption}

\begin{figure}[t!]
    \includegraphics[width=\columnwidth]{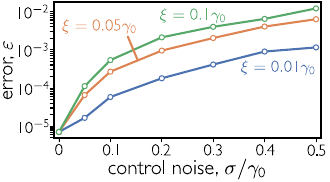}
    \caption{Error in the photon emission from a two-emitter system with $s=\lambda_0/4$ and $d=\lambda_0$ and noisy control over the local detuning. The error is defined as the $L^2$ distance between the emitted photon profile and a target Gaussian photon with $T=8 / \gamma_0$ and $\tau = 1.25 / \gamma_0$. The detuning of each emitter is subject to a Gaussian noise with standard deviation $\sigma$, and a new random sample is drawn after time $\xi$.
}
    \label{fig: error_emission}
\end{figure}

In this Appendix, we show that storage and emission protocols are robust to noise in the applied temporal detuning sequence. We consider Gaussian noise with standard deviation $\sigma$ that provides a constant error during a time bin of length $\xi$. We further define the error $\epsilon$ as the $L^2$ distance between the resulting photon $\mathcal{P}(t)$ and the target Gaussian photon $\mathcal{P}_G(t)$, that is, $\epsilon = \int_0^T dt |\mathcal{P}(t) - \mathcal{P}_G(t)|^2$. In Fig.~\ref{fig: error_emission}, we plot the error for the emission of a single photon from a two-emitter system. We obtain $\epsilon \sim 10^{-5}$ in the absence of control noise. The error is extremely robust if the noise changes fast compared to the emitter timescales (\ie small $\xi$), as it largely averages out. While slower noise causes larger errors, the protocol remains remarkably robust.

\bibliography{bibliography}

@article{Fernandez22,
	author = {Fern\'andez-Fern\'andez, D. and Gonz\'alez-Tudela, A.},
	doi = {10.1103/PhysRevLett.128.113601},
	issue = {11},
	journal = {Phys. Rev. Lett.},
	month = {Mar},
	numpages = {6},
	pages = {113601},
	publisher = {American Physical Society},
	title = {Tunable Directional Emission and Collective Dissipation with Quantum Metasurfaces},
	url = {https://link.aps.org/doi/10.1103/PhysRevLett.128.113601},
	volume = {128},
	year = {2022}}

@article{Almanakly25,
	author = {Almanakly, Aziza and Yankelevich, Beatriz and Hays, Max and Kannan, Bharath and Assouly, R{\'e}ouven and Greene, Alex and Gingras, Michael and Niedzielski, Bethany M. and Stickler, Hannah and Schwartz, Mollie E. and Serniak, Kyle and Wang, Joel {\^I}-j. and Orlando, Terry P. and Gustavsson, Simon and Grover, Jeffrey A. and Oliver, William D.},
	date = {2025/05/01},
	doi = {10.1038/s41567-025-02811-1},
	id = {Almanakly2025},
	isbn = {1745-2481},
	journal = {Nat. Phys.},
	number = {5},
	pages = {825--830},
	title = {Deterministic remote entanglement using a chiral quantum interconnect},
	url = {https://doi.org/10.1038/s41567-025-02811-1},
	volume = {21},
	year = {2025}}

@article{OSullivan25,
	author = {O'Sullivan, James and Reuer, Kevin and Grigorev, Aleksandr and Dai, Xi and Hern{\'a}ndez-Ant{\'o}n, Alonso and Mu{\~n}oz-Arias, Manuel H. and Hellings, Christoph and Flasby, Alexander and Colao Zanuz, Dante and Besse, Jean-Claude and Blais, Alexandre and Malz, Daniel and Eichler, Christopher and Wallraff, Andreas},
	date = {2025/07/01},
	doi = {10.1038/s41467-025-60472-3},
	id = {O'Sullivan2025},
	isbn = {2041-1723},
	journal = {Nat. Commun.},
	number = {1},
	pages = {5505},
	title = {Deterministic generation of two-dimensional multi-photon cluster states},
	url = {https://doi.org/10.1038/s41467-025-60472-3},
	volume = {16},
	year = {2025}}

@article{Hoi15,
	author = {Hoi, I. -C. and Kockum, A. F. and Tornberg, L. and Pourkabirian, A. and Johansson, G. and Delsing, P. and Wilson, C. M.},
	date = {2015/12/01},
	doi = {10.1038/nphys3484},
	id = {Hoi2015},
	isbn = {1745-2481},
	journal = {Nat. Phys.},
	number = {12},
	pages = {1045--1049},
	title = {Probing the quantum vacuum with an artificial atom in front of a mirror},
	url = {https://doi.org/10.1038/nphys3484},
	volume = {11},
	year = {2015}}

@article{Facchinetti18,
	author = {Facchinetti, G. and Ruostekoski, J.},
	doi = {10.1103/PhysRevA.97.023833},
	issue = {2},
	journal = {Phys. Rev. A},
	month = {Feb},
	numpages = {12},
	pages = {023833},
	publisher = {American Physical Society},
	title = {Interaction of light with planar lattices of atoms: Reflection, transmission, and cooperative magnetometry},
	url = {https://link.aps.org/doi/10.1103/PhysRevA.97.023833},
	volume = {97},
	year = {2018}}

@article{Wang26,
	author = {Wang, Xin and Liao, Zeyang},
	doi = {10.1103/4crz-846z},
	issue = {2},
	journal = {Phys. Rev. Appl.},
	month = {Aug},
	numpages = {9},
	pages = {024008},
	publisher = {American Physical Society},
	title = {Super-{H}eisenberg-limited sensing via collective subradiance in waveguide quantum electrodynamics},
	url = {https://link.aps.org/doi/10.1103/4crz-846z},
	volume = {26},
	year = {2026}}

@article{Zafra26,
	author = {Zafra-Bono, Diego and Rubies-Bigorda, Oriol and Yelin, Susanne F.},
	doi = {10.1103/rprq-bp61},
	issue = {1},
	journal = {Phys. Rev. A},
	month = {Jul},
	numpages = {14},
	pages = {013704},
	publisher = {American Physical Society},
	title = {Subradiant collective states for precision sensing via transmission spectra},
	url = {https://link.aps.org/doi/10.1103/rprq-bp61},
	volume = {114},
	year = {2026}}

@article{Ballantine21,
	author = {Ballantine, K. E. and Ruostekoski, J.},
	doi = {10.1103/PRXQuantum.2.040362},
	issue = {4},
	journal = {PRX Quantum},
	month = {Dec},
	numpages = {14},
	pages = {040362},
	publisher = {American Physical Society},
	title = {Quantum Single-Photon Control, Storage, and Entanglement Generation with Planar Atomic Arrays},
	url = {https://link.aps.org/doi/10.1103/PRXQuantum.2.040362},
	volume = {2},
	year = {2021}}

@article{Palma96,
	author = {Palma, G. Massimo and Suominen, Kalle-{A}ntti and Ekert, Artur},
	doi = {10.1098/rspa.1996.0029},
	isbn = {1364-5021},
	journal = {Proc. R. Soc. A: Math. Phys. Eng. Sci.},
	journal1 = {Proc. R. Soc. A},
	month = {8/13/2026},
	number = {1946},
	pages = {567--584},
	title = {Quantum computers and dissipation},
	url = {https://doi.org/10.1098/rspa.1996.0029},
	volume = {452},
	year = {1996},
	year1 = {1996/12/31}}

@article{Douglas26arxiv,
	author = {Alec Douglas and Lin Su and Michal Szurek and Robin Groth and Sandra Brandstetter and Ognjen Markovi\'{c} and Oriol Rubies-Bigorda and Stefan Ostermann and Susanne F. Yelin and Markus Greiner},
	journal = {arXiv:2604.11795},
	title = {Many-Body Super- and Subradiance in Ordered Atomic Arrays},
	year = {2026}}

@article{Beige00,
	author = {Beige, Almut and Braun, Daniel and Tregenna, Ben and Knight, Peter L.},
	doi = {10.1103/PhysRevLett.85.1762},
	issue = {8},
	journal = {Phys. Rev. Lett.},
	month = {Aug},
	numpages = {0},
	pages = {1762--1765},
	publisher = {American Physical Society},
	title = {Quantum Computing Using Dissipation to Remain in a Decoherence-Free Subspace},
	url = {https://link.aps.org/doi/10.1103/PhysRevLett.85.1762},
	volume = {85},
	year = {2000}}

@article{Ocola24,
	author = {Ocola, Paloma L. and Dimitrova, Ivana and Grinkemeyer, Brandon and Guardado-Sanchez, Elmer and \DJ{}or\dj{}evi\ifmmode \acute{c}\else \'{c}\fi{}, Tamara and Samutpraphoot, Polnop and Vuleti\ifmmode \acute{c}\else \'{c}\fi{}, Vladan and Lukin, Mikhail D.},
	doi = {10.1103/PhysRevLett.132.113601},
	issue = {11},
	journal = {Phys. Rev. Lett.},
	month = {Mar},
	numpages = {6},
	pages = {113601},
	publisher = {American Physical Society},
	title = {Control and Entanglement of Individual {R}ydberg Atoms near a Nanoscale Device},
	url = {https://link.aps.org/doi/10.1103/PhysRevLett.132.113601},
	volume = {132},
	year = {2024}}

@article{Shah24,
	author = {Shah, Freya and Patti, Taylor L. and Rubies-Bigorda, Oriol and Yelin, Susanne F.},
	doi = {10.1103/PhysRevA.109.012613},
	issue = {1},
	journal = {Phys. Rev. A},
	month = {Jan},
	numpages = {16},
	pages = {012613},
	publisher = {American Physical Society},
	title = {Quantum computing with subwavelength atomic arrays},
	url = {https://link.aps.org/doi/10.1103/PhysRevA.109.012613},
	volume = {109},
	year = {2024}}

@article{Facchi02,
	author = {Facchi, P. and Pascazio, S.},
	doi = {10.1103/PhysRevLett.89.080401},
	issue = {8},
	journal = {Phys. Rev. Lett.},
	month = {Aug},
	numpages = {4},
	pages = {080401},
	publisher = {American Physical Society},
	title = {Quantum {Z}eno Subspaces},
	url = {https://link.aps.org/doi/10.1103/PhysRevLett.89.080401},
	volume = {89},
	year = {2002}}

@article{Wiegand20,
	author = {Wiegand, E. and Rousseaux, B. and Johansson, G.},
	doi = {10.1103/PhysRevA.101.033801},
	issue = {3},
	journal = {Phys. Rev. A},
	month = {Mar},
	numpages = {13},
	pages = {033801},
	publisher = {American Physical Society},
	title = {Semiclassical analysis of dark-state transient dynamics in waveguide circuit {QED}},
	url = {https://link.aps.org/doi/10.1103/PhysRevA.101.033801},
	volume = {101},
	year = {2020}}

@article{Du26arxiv,
	archiveprefix = {arXiv},
	author = {Botao Du and Qihao Guo and Ruichao Ma},
	eprint = {2605.12442},
	journal = {arXiv:2605.12442},
	primaryclass = {cond-mat.quant-gas},
	title = {Programmable Superradiance in an Interacting Qubit Array},
	url = {https://arxiv.org/abs/2605.12442},
	year = {2026}}

@article{Lalumiere13,
	author = {Lalumi\`ere, Kevin and Sanders, Barry C. and van Loo, A. F. and Fedorov, A. and Wallraff, A. and Blais, A.},
	doi = {10.1103/PhysRevA.88.043806},
	issue = {4},
	journal = {Phys. Rev. A},
	month = {Oct},
	numpages = {15},
	pages = {043806},
	publisher = {American Physical Society},
	title = {Input-output theory for waveguide {QED} with an ensemble of inhomogeneous atoms},
	url = {https://link.aps.org/doi/10.1103/PhysRevA.88.043806},
	volume = {88},
	year = {2013}}

@article{Zanardi97,
	author = {Zanardi, P. and Rasetti, M.},
	doi = {10.1103/PhysRevLett.79.3306},
	issue = {17},
	journal = {Phys. Rev. Lett.},
	month = {Oct},
	numpages = {0},
	pages = {3306--3309},
	publisher = {American Physical Society},
	title = {Noiseless Quantum Codes},
	url = {https://link.aps.org/doi/10.1103/PhysRevLett.79.3306},
	volume = {79},
	year = {1997}}

@article{Knaut24,
	author = {Knaut, C. M. and Suleymanzade, A. and Wei, Y. -C. and Assumpcao, D. R. and Stas, P. -J. and Huan, Y. Q. and Machielse, B. and Knall, E. N. and Sutula, M. and Baranes, G. and Sinclair, N. and De-Eknamkul, C. and Levonian, D. S. and Bhaskar, M. K. and Park, H. and Lon{\v c}ar, M. and Lukin, M. D.},
	date = {2024/05/01},
	doi = {10.1038/s41586-024-07252-z},
	id = {Knaut2024},
	isbn = {1476-4687},
	journal = {Nature},
	number = {8012},
	pages = {573--578},
	title = {Entanglement of nanophotonic quantum memory nodes in a telecom network},
	url = {https://doi.org/10.1038/s41586-024-07252-z},
	volume = {629},
	year = {2024}}

@article{Fayard21,
	author = {Fayard, N. and Henriet, L. and Asenjo-Garcia, A. and Chang, D. E.},
	doi = {10.1103/PhysRevResearch.3.033233},
	issue = {3},
	journal = {Phys. Rev. Res.},
	month = {Sep},
	numpages = {15},
	pages = {033233},
	publisher = {American Physical Society},
	title = {Many-body localization in waveguide quantum electrodynamics},
	url = {https://link.aps.org/doi/10.1103/PhysRevResearch.3.033233},
	volume = {3},
	year = {2021}}

@article{Rusconi21,
	author = {Rusconi, Cosimo C. and Shi, Tao and Cirac, J. Ignacio},
	doi = {10.1103/PhysRevA.104.033718},
	issue = {3},
	journal = {Phys. Rev. A},
	month = {Sep},
	numpages = {22},
	pages = {033718},
	publisher = {American Physical Society},
	title = {Exploiting the photonic nonlinearity of free-space subwavelength arrays of atoms},
	url = {https://link.aps.org/doi/10.1103/PhysRevA.104.033718},
	volume = {104},
	year = {2021}}

@article{Holzinger22,
	author = {Holzinger, R. and Guti\'errez-J\'auregui, R. and H\"onigl-Decrinis, T. and Kirchmair, G. and Asenjo-Garcia, A. and Ritsch, H.},
	doi = {10.1103/PhysRevLett.129.253601},
	issue = {25},
	journal = {Phys. Rev. Lett.},
	month = {Dec},
	numpages = {6},
	pages = {253601},
	publisher = {American Physical Society},
	title = {Control of Localized Single- and Many-Body Dark States in Waveguide {QED}},
	url = {https://link.aps.org/doi/10.1103/PhysRevLett.129.253601},
	volume = {129},
	year = {2022}}

@article{Rubies25,
	author = {Rubies-Bigorda, Oriol and Masson, Stuart J. and Yelin, Susanne F. and Asenjo-Garcia, Ana},
	doi = {10.1103/PhysRevLett.134.213603},
	issue = {21},
	journal = {Phys. Rev. Lett.},
	month = {May},
	numpages = {7},
	pages = {213603},
	publisher = {American Physical Society},
	title = {Deterministic Generation of Photonic Entangled States Using Decoherence-Free Subspaces},
	url = {https://link.aps.org/doi/10.1103/PhysRevLett.134.213603},
	volume = {134},
	year = {2025}}

@article{Ferreira24,
	author = {Ferreira, Vinicius S. and Kim, Gihwan and Butler, Andreas and Pichler, Hannes and Painter, Oskar},
	date = {2024/05/01},
	doi = {10.1038/s41567-024-02408-0},
	id = {Ferreira2024},
	isbn = {1745-2481},
	journal = {Nat. Phys.},
	number = {5},
	pages = {865--870},
	title = {Deterministic generation of multidimensional photonic cluster states with a single quantum emitter},
	url = {https://doi.org/10.1038/s41567-024-02408-0},
	volume = {20},
	year = {2024}}

@article{Kannan23,
	author = {Kannan, Bharath and Almanakly, Aziza and Sung, Youngkyu and Di Paolo, Agustin and Rower, David A. and Braum{\"u}ller, Jochen and Melville, Alexander and Niedzielski, Bethany M. and Karamlou, Amir and Serniak, Kyle and Veps{\"a}l{\"a}inen, Antti and Schwartz, Mollie E. and Yoder, Jonilyn L. and Winik, Roni and Wang, Joel I-Jan and Orlando, Terry P. and Gustavsson, Simon and Grover, Jeffrey A. and Oliver, William D.},
	date = {2023/03/01},
	doi = {10.1038/s41567-022-01869-5},
	id = {Kannan2023},
	isbn = {1745-2481},
	journal = {Nat. Phys.},
	number = {3},
	pages = {394--400},
	title = {On-demand directional microwave photon emission using waveguide quantum electrodynamics},
	url = {https://doi.org/10.1038/s41567-022-01869-5},
	volume = {19},
	year = {2023}}

@article{Zanner22,
	author = {Zanner, Maximilian and Orell, Tuure and Schneider, Christian M. F. and Albert, Romain and Oleschko, Stefan and Juan, Mathieu L. and Silveri, Matti and Kirchmair, Gerhard},
	date = {2022/05/01},
	doi = {10.1038/s41567-022-01527-w},
	id = {Zanner2022},
	isbn = {1745-2481},
	journal = {Nat. Phys.},
	number = {5},
	pages = {538--543},
	title = {Coherent control of a multi-qubit dark state in waveguide quantum electrodynamics},
	url = {https://doi.org/10.1038/s41567-022-01527-w},
	volume = {18},
	year = {2022}}

@article{Zhenjie23,
	author = {Yan, Zhenjie and Ho, Jacquelyn and Lu, Yue-Hui and Masson, Stuart J. and Asenjo-Garcia, Ana and Stamper-Kurn, Dan M.},
	doi = {10.1103/PhysRevLett.131.253603},
	issue = {25},
	journal = {Phys. Rev. Lett.},
	month = {Dec},
	numpages = {7},
	pages = {253603},
	publisher = {American Physical Society},
	title = {Superradiant and Subradiant Cavity Scattering by Atom Arrays},
	url = {https://link.aps.org/doi/10.1103/PhysRevLett.131.253603},
	volume = {131},
	year = {2023}}

@article{Tiranov23,
	author = {Tiranov, Alexey and Angelopoulou, Vasiliki and van Diepen, Cornelis Jacobus and Schrinski, Bj{\"o}rn and Sandberg, Oliver August Dall'Alba and Wang, Ying and Midolo, Leonardo and Scholz, Sven and Wieck, Andreas Dirk and Ludwig, Arne and S{\o}rensen, Anders S{\o}ndberg and Lodahl, Peter},
	date = {2023/01/27},
	doi = {10.1126/science.ade9324},
	journal = {Science},
	journal1 = {Science},
	journal2 = {Science},
	month = {2023/11/27},
	n2 = {Photon emission is the hallmark of light-matter interaction and the foundation of photonic quantum science, enabling advanced sources for quantum communication and computing. Although single-emitter radiation can be tailored by the photonic environment, the introduction of multiple emitters extends this picture. A fundamental challenge, however, is that the radiative dipole-dipole coupling rapidly decays with spatial separation, typically within a fraction of the optical wavelength. We realize distant dipole-dipole radiative coupling with pairs of solid-state optical quantum emitters embedded in a nanophotonic waveguide. We dynamically probe the collective response and identify both super- and subradiant emission as well as means to control the dynamics by proper excitation techniques. Our work constitutes a foundational step toward multiemitter applications for scalable quantum-information processing. The merger of nanophotonics with quantum optics has provided a platform for the development of deterministic single-photon sources and sources of entangled photons. Tiranov et al. show that multiple quantum emitters (two or three quantum dots) can be coupled through a photonic crystal waveguide. The waveguide helps to overcome the typically short-range nature of the dipole-dipole interactions and allows the signature properties of such a coupled system, subradiant and superradiant emission, to be observed. Control of such processes could provide an enabling step for scaling up deterministic solid-state photon-emitter interfaces and multiphoton-entangled sources for applications in quantum information processing. ?ISO A photonic crystal waveguide enables distant coupling between multiple solid-state optical quantum emitters.},
	number = {6630},
	pages = {389--393},
	publisher = {American Association for the Advancement of Science},
	title = {Collective super- and subradiant dynamics between distant optical quantum emitters},
	type = {doi: 10.1126/science.ade9324},
	url = {https://doi.org/10.1126/science.ade9324},
	volume = {379},
	year = {2023},
	year1 = {2023}}

@article{Rubies23PRA,
	author = {Rubies-Bigorda, Oriol and Ostermann, Stefan and Yelin, Susanne F.},
	doi = {10.1103/PhysRevA.107.L051701},
	issue = {5},
	journal = {Phys. Rev. A},
	month = {May},
	numpages = {6},
	pages = {L051701},
	publisher = {American Physical Society},
	title = {Dynamic population of multiexcitation subradiant states in incoherently excited atomic arrays},
	url = {https://link.aps.org/doi/10.1103/PhysRevA.107.L051701},
	volume = {107},
	year = {2023}}

@article{Rubies22PRR,
	author = {Rubies-Bigorda, Oriol and Walther, Valentin and Patti, Taylor L. and Yelin, Susanne F.},
	doi = {10.1103/PhysRevResearch.4.013110},
	issue = {1},
	journal = {Phys. Rev. Research},
	month = {Feb},
	numpages = {17},
	pages = {013110},
	publisher = {American Physical Society},
	title = {Photon control and coherent interactions via lattice dark states in atomic arrays},
	url = {https://link.aps.org/doi/10.1103/PhysRevResearch.4.013110},
	volume = {4},
	year = {2022}}

@article{Santos22,
	author = {Santos, Alan C. and Cidrim, Andr\'e and Villas-Boas, Celso Jorge and Kaiser, Robin and Bachelard, Romain},
	doi = {10.1103/PhysRevA.105.053715},
	issue = {5},
	journal = {Phys. Rev. A},
	month = {May},
	numpages = {6},
	pages = {053715},
	publisher = {American Physical Society},
	title = {Generating long-lived entangled states with free-space collective spontaneous emission},
	url = {https://link.aps.org/doi/10.1103/PhysRevA.105.053715},
	volume = {105},
	year = {2022}}

@article{CastellsGraells21,
	author = {Castells-Graells, David and Malz, Daniel and Rusconi, Cosimo C. and Cirac, J. Ignacio},
	doi = {10.1103/PhysRevA.104.063707},
	issue = {6},
	journal = {Phys. Rev. A},
	month = {Dec},
	numpages = {16},
	pages = {063707},
	publisher = {American Physical Society},
	title = {Atomic waveguide QED with atomic dimers},
	url = {https://link.aps.org/doi/10.1103/PhysRevA.104.063707},
	volume = {104},
	year = {2021}}

@article{Paulisch16,
	author = {V Paulisch and H J Kimble and A Gonz{\'{a}}lez-Tudela},
	doi = {10.1088/1367-2630/18/4/043041},
	journal = {New J. Phys.},
	month = {apr},
	number = {4},
	pages = {043041},
	publisher = {{IOP} Publishing},
	title = {Universal quantum computation in waveguide {QED} using decoherence free subspaces},
	url = {https://doi.org/10.1088/1367-2630/18/4/043041},
	volume = {18},
	year = 2016}

@article{Cidrim20,
	author = {Cidrim, A. and do Espirito Santo, T. S. and Schachenmayer, J. and Kaiser, R. and Bachelard, R.},
	doi = {10.1103/PhysRevLett.125.073601},
	issue = {7},
	journal = {Phys. Rev. Lett.},
	month = {Aug},
	numpages = {6},
	pages = {073601},
	publisher = {American Physical Society},
	title = {Photon Blockade with Ground-State Neutral Atoms},
	url = {https://link.aps.org/doi/10.1103/PhysRevLett.125.073601},
	volume = {125},
	year = {2020}}

@article{Brechtelsbauer21,
	author = {Brechtelsbauer, Katharina and Malz, Daniel},
	doi = {10.1103/PhysRevA.104.013701},
	issue = {1},
	journal = {Phys. Rev. A},
	month = {Jul},
	numpages = {13},
	pages = {013701},
	publisher = {American Physical Society},
	title = {Quantum simulation with fully coherent dipole-dipole interactions mediated by three-dimensional subwavelength atomic arrays},
	url = {https://link.aps.org/doi/10.1103/PhysRevA.104.013701},
	volume = {104},
	year = {2021}}

@article{Patti21,
	author = {Patti, Taylor L. and Wild, Dominik S. and Shahmoon, Ephraim and Lukin, Mikhail D. and Yelin, Susanne F.},
	doi = {10.1103/PhysRevLett.126.223602},
	issue = {22},
	journal = {Phys. Rev. Lett.},
	month = {Jun},
	numpages = {6},
	pages = {223602},
	publisher = {American Physical Society},
	title = {Controlling Interactions between Quantum Emitters Using Atom Arrays},
	url = {https://link.aps.org/doi/10.1103/PhysRevLett.126.223602},
	volume = {126},
	year = {2021}}

@article{Ferioli21PRX,
	author = {Ferioli, Giovanni and Glicenstein, Antoine and Henriet, Loic and Ferrier-Barbut, Igor and Browaeys, Antoine},
	doi = {10.1103/PhysRevX.11.021031},
	issue = {2},
	journal = {Phys. Rev. X},
	month = {May},
	numpages = {12},
	pages = {021031},
	publisher = {American Physical Society},
	title = {Storage and Release of Subradiant Excitations in a Dense Atomic Cloud},
	url = {https://link.aps.org/doi/10.1103/PhysRevX.11.021031},
	volume = {11},
	year = {2021}}

@article{Masson20PRR,
	author = {Masson, Stuart J. and Asenjo-Garcia, Ana},
	doi = {10.1103/PhysRevResearch.2.043213},
	issue = {4},
	journal = {Phys. Rev. Research},
	month = {Nov},
	numpages = {16},
	pages = {043213},
	publisher = {American Physical Society},
	title = {Atomic-waveguide quantum electrodynamics},
	volume = {2},
	year = {2020}}

@book{M}

@article{Williamson20PRL,
	author = {Williamson, L. A. and Borgh, M. O. and Ruostekoski, J.},
	doi = {10.1103/PhysRevLett.125.073602},
	issue = {7},
	journal = {Phys. Rev. Lett.},
	month = {Aug},
	numpages = {7},
	pages = {073602},
	publisher = {American Physical Society},
	title = {Superatom Picture of Collective Nonclassical Light Emission and Dipole Blockade in Atom Arrays},
	url = {https://link.aps.org/doi/10.1103/PhysRevLett.125.073602},
	volume = {125},
	year = {2020}}

@article{Rui20,
	author = {Rui, Jun and Wei, David and Rubio-Abadal, Antonio and Hollerith, Simon and Zeiher, Johannes and Stamper-Kurn, Dan M. and Gross, Christian and Bloch, Immanuel},
	da = {2020/07/01},
	doi = {10.1038/s41586-020-2463-x},
	id = {Rui2020},
	isbn = {1476-4687},
	journal = {Nature},
	number = {7816},
	pages = {369--374},
	title = {A subradiant optical mirror formed by a single structured atomic layer},
	ty = {JOUR},
	volume = {583},
	year = {2020}}

@article{Solano17,
	author = {Solano, P. and Barberis-Blostein, P. and Fatemi, F. K. and Orozco, L. A. and Rolston, S. L.},
	da = {2017/11/30},
	doi = {10.1038/s41467-017-01994-3},
	id = {Solano2017},
	isbn = {2041-1723},
	journal = {Nat. Commun.},
	number = {1},
	pages = {1857},
	title = {Super-radiance reveals infinite-range dipole interactions through a nanofiber},
	ty = {JOUR},
	url = {https://doi.org/10.1038/s41467-017-01994-3},
	volume = {8},
	year = {2017}}

@article{Guerin16,
	author = {Guerin, William and Ara\'ujo, Michelle O. and Kaiser, Robin},
	doi = {10.1103/PhysRevLett.116.083601},
	issue = {8},
	journal = {Phys. Rev. Lett.},
	month = {Feb},
	numpages = {5},
	pages = {083601},
	publisher = {American Physical Society},
	title = {Subradiance in a Large Cloud of Cold Atoms},
	url = {https://link.aps.org/doi/10.1103/PhysRevLett.116.083601},
	volume = {116},
	year = {2016}}

@article{Gross82,
	author = {Gross, M. and Haroche, S.},
	da = {1982/12/01/},
	doi = {https://doi.org/10.1016/0370-1573(82)90102-8},
	isbn = {0370-1573},
	journal = {Phys. Rep.},
	number = {5},
	pages = {301--396},
	title = {Superradiance: An essay on the theory of collective spontaneous emission},
	ty = {JOUR},
	volume = {93},
	year = {1982}}

@article{Henriet19,
	author = {Henriet, Lo\"{\i}c and Douglas, James S. and Chang, Darrick E. and Albrecht, Andreas},
	doi = {10.1103/PhysRevA.99.023802},
	issue = {2},
	journal = {Phys. Rev. A},
	month = {Feb},
	numpages = {20},
	pages = {023802},
	publisher = {American Physical Society},
	title = {Critical open-system dynamics in a one-dimensional optical-lattice clock},
	url = {https://link.aps.org/doi/10.1103/PhysRevA.99.023802},
	volume = {99},
	year = {2019}}

@article{Zhang19,
	author = {Zhang, Yu-Xiang and M\o{}lmer, Klaus},
	doi = {10.1103/PhysRevLett.122.203605},
	issue = {20},
	journal = {Phys. Rev. Lett.},
	month = {May},
	numpages = {5},
	pages = {203605},
	publisher = {American Physical Society},
	title = {Theory of Subradiant States of a One-Dimensional Two-Level Atom Chain},
	url = {https://link.aps.org/doi/10.1103/PhysRevLett.122.203605},
	volume = {122},
	year = {2019}}

@article{Goban15,
	author = {Goban, A. and Hung, C.-L. and Hood, J. D. and Yu, S.-P. and Muniz, J. A. and Painter, O. and Kimble, H. J.},
	doi = {10.1103/PhysRevLett.115.063601},
	issue = {6},
	journal = {Phys. Rev. Lett.},
	month = {Aug},
	numpages = {5},
	pages = {063601},
	publisher = {American Physical Society},
	title = {Superradiance for Atoms Trapped along a Photonic Crystal Waveguide},
	url = {https://link.aps.org/doi/10.1103/PhysRevLett.115.063601},
	volume = {115},
	year = {2015}}

@article{Cirac97,
	author = {Cirac, J. I. and Zoller, P. and Kimble, H. J. and Mabuchi, H.},
	doi = {10.1103/PhysRevLett.78.3221},
	issue = {16},
	journal = {Phys. Rev. Lett.},
	month = {Apr},
	numpages = {0},
	pages = {3221--3224},
	publisher = {American Physical Society},
	title = {Quantum State Transfer and Entanglement Distribution among Distant Nodes in a Quantum Network},
	url = {https://link.aps.org/doi/10.1103/PhysRevLett.78.3221},
	volume = {78},
	year = {1997}}

@article{Mirhosseini19,
	author = {Mirhosseini, Mohammad and Kim, Eunjong and Zhang, Xueyue and Sipahigil, Alp and Dieterle, Paul B. and Keller, Andrew J. and Asenjo-Garcia, Ana and Chang, Darrick E. and Painter, Oskar},
	da = {2019/05/01},
	doi = {10.1038/s41586-019-1196-1},
	id = {Mirhosseini2019},
	isbn = {1476-4687},
	journal = {Nature},
	number = {7758},
	pages = {692--697},
	title = {Cavity quantum electrodynamics with atom-like mirrors},
	ty = {JOUR},
	url = {https://doi.org/10.1038/s41586-019-1196-1},
	volume = {569},
	year = {2019}}

@article{Asenjo17PRX,
	author = {Asenjo-Garcia, A. and Moreno-Cardoner, M. and Albrecht, A. and Kimble, H. J. and Chang, D. E.},
	doi = {10.1103/PhysRevX.7.031024},
	issue = {3},
	journal = {Phys. Rev. X},
	month = {Aug},
	numpages = {36},
	pages = {031024},
	publisher = {American Physical Society},
	title = {Exponential Improvement in Photon Storage Fidelities Using Subradiance and ``Selective Radiance'' in Atomic Arrays},
	volume = {7},
	year = {2017}}

@article{Manzoni18,
	author = {M T Manzoni and M Moreno-Cardoner and A Asenjo-Garcia and J V Porto and A V Gorshkov and D E Chang},
	doi = {10.1088/1367-2630/aadb74},
	journal = {New J. Phys.},
	month = {aug},
	number = {8},
	pages = {083048},
	publisher = {{IOP} Publishing},
	title = {Optimization of photon storage fidelity in ordered atomic arrays},
	url = {https://doi.org/10.1088%2F1367-2630%2Faadb74},
	volume = {20},
	year = 2018}

@article{Dung02,
	author = {Dung, Ho Trung and Kn\"oll, Ludwig and Welsch, Dirk-Gunnar},
	doi = {10.1103/PhysRevA.66.063810},
	issue = {6},
	journal = {Phys. Rev. A},
	month = {Dec},
	numpages = {16},
	pages = {063810},
	publisher = {American Physical Society},
	title = {Resonant dipole-dipole interaction in the presence of dispersing and absorbing surroundings},
	url = {https://link.aps.org/doi/10.1103/PhysRevA.66.063810},
	volume = {66},
	year = {2002}}

@article{Lidar98,
	author = {Lidar, D. A. and Chuang, I. L. and Whaley, K. B.},
	doi = {10.1103/PhysRevLett.81.2594},
	issue = {12},
	journal = {Phys. Rev. Lett.},
	month = {Sep},
	numpages = {0},
	pages = {2594--2597},
	publisher = {American Physical Society},
	title = {Decoherence-Free Subspaces for Quantum Computation},
	url = {https://link.aps.org/doi/10.1103/PhysRevLett.81.2594},
	volume = {81},
	year = {1998}}

@article{Duan01,
	author = {Duan, L. -M. and Lukin, M. D. and Cirac, J. I. and Zoller, P.},
	date = {2001/11/22/print},
	day = {22},
	isbn = {0028-0836},
	journal = {Nature},
	l3 = {http://www.nature.com/nature/journal/v414/n6862/suppinfo/414413a0_S1.html},
	m3 = {10.1038/35106500},
	month = {11},
	number = {6862},
	pages = {413--418},
	publisher = {Macmillian Magazines Ltd.},
	title = {Long-distance quantum communication with atomic ensembles and linear optics},
	ty = {JOUR},
	url = {http://dx.doi.org/10.1038/35106500},
	volume = {414},
	year = {2001}}

@article{Dicke54,
	author = {Dicke, R. H.},
	doi = {10.1103/PhysRev.93.99},
	issue = {1},
	journal = {Phys. Rev.},
	month = {Jan},
	numpages = {0},
	pages = {99--110},
	publisher = {American Physical Society},
	title = {Coherence in Spontaneous Radiation Processes},
	volume = {93},
	year = {1954}}

@article{Goban14,
	author = {Goban, A. and Hung, C. -L. and Yu, S. -P. and Hood, J. D. and Muniz, J. A. and Lee, J. H. and Martin, M. J. and McClung, A. C. and Choi, K. S. and Chang, D. E. and Painter, O. and Kimble, H. J.},
	date = {2014/05/08/online},
	day = {08},
	journal = {Nat. Commun.},
	l3 = {10.1038/ncomms4808},
	m3 = {Article},
	month = {05},
	number = {3808},
	publisher = {Nature Publishing Group, a division of Macmillan Publishers Limited. All Rights Reserved.},
	title = {Atom--light interactions in photonic crystals},
	ty = {JOUR},
	url = {http://dx.doi.org/10.1038/ncomms4808},
	volume = {5},
	year = {2014}}
\end{document}